# Euclidean Approach to Green-Wave Theory Applied to Traffic Signal Networks

Melvin H. Friedman[1], Brian L. Mark[1], and Nathan H. Gartner[2]

[1] Dept. of Electrical and Computer Engineering, George Mason University, 4400 University Dr, Fairfax, VA 22030, United States.
[2] Dept. of Civil Engineering, Ariel University, Ariel, Israel

## Abstract

Travel on long arterials with signalized intersections can be inefficient if not coordinated properly. As the number of signals increases, coordination becomes more challenging and traditional progression schemes tend to break down. Long progressions save travel time and fuel, reduce pollution and traffic accidents by providing a smoother flow of traffic. This paper introduces a green-wave theory that can be applied to a network of intersecting arterial roads. It enables uninterrupted flow on arbitrary long signalized arterials using a Road-to-Traveler-Feedback Device. The approach is modelled after Euclid. We define concepts such as RGW-roads (roads where vehicles traveling at recommended speed make all traffic signals), green-arrows (representing vehicle platoons), real nodes (representing signalized intersections where RGW-roads intersect) and virtual nodes, green-wave speed, blocks, etc. – the analogue of Euclid's points, lines, parallel lines. We postulate green-arrow laws of motion – the analogue of Euclid's postulates. We then use geometric reasoning to deduce results: green-arrow lengths have a maximum value, are restricted to discrete lengths, and green-arrow laws of motion imply that select existing arterial roads can be converted to RGW-roads. The signal timings and offsets that are produced have been shown to be effective using a simulation model developed previously called RGW-SIM.

# 1. Introduction

Coordination of traffic lights along arterial streets is essential for obtaining continuous flow with minimum interruptions and maximum utilization of the road space. A recently published paper [1] demonstrated how to achieve uninterrupted flow, with maximum bandwidth and throughput, on an existing 17 km two-way arterial road (Telegraph Road, Alexandria, VA USA). This accomplishment was made possible by the *Road-to-Traveler-Feedback-Device* (RTFD) which advises motorists how fast they should travel to stay within the progression band. A one-page summary of the RTFD is available in [2]. A patent application has been submitted to the US Patent Office describing an implementation of this device [3].

Reference [1] focused on how to get uninterrupted flow with maximum throughput on a single two-way suburban road. This paper extends the methodology to an arbitrary network of two-way roads topologically equivalent to a Cartesian network of roads. The challenge is to provide simultaneous progressions on all the arterials of the network, which implies that progression bands on intersecting arterials cannot overlap in time at any junction of the network. Development of progressions in networks is subject to the so-called "loop constraint" that governs synchronized signal networks. It states that "for any closed loops of the network consisting of more than two links, the summation of *internode* and *intranode offsets* around the loop of intersecting arterials must be an integer multiple of the cycle time" [4]. A procedure for determining a set of independent loops in a network is described in [5]. Existing approaches to the network synchronization problem are based primarily on mathematical programming techniques [4][6][7][8][9][10]. Commercial approaches to network signal synchronization are based on delay-and -stops minimization. Notable delay-and-stops-based models are TRANSYT and SYNCHRO. TRANSYT (Traffic Network Study Tool) was originally developed by Robertson at the UK Transport Road Research Laboratory [11] and has since been revised multiple times, now in its 16th generation [12]. SYNCHRO is a more recently developed model distributed by Cubic Transportation Systems in the US [13]. Both models use iterative search procedures to optimize traffic signal timings based on simulated traffic conditions. They aim to minimize overall delay and stop frequency and maximize progression opportunities along corridors. Both have extensive graphic capabilities to aid in the design process and are widely used in practice. Recent approaches attempt to provide dedicated progression bands for major origin-destination flows along arterial routes [14][15]. Arsava et al [16] extends this approach to address the OD based traffic signal coordination problem in multi-arterial grid networks. The extended model can create separate progression bands for each major OD stream in the network. Similar approaches have been introduced for the combined control of vehicular traffic and transit vehicles [17][18][19]. These approaches attempt to thread bus and tram progressions through signalized intersections simultaneously with vehicular progressions.

All these models attempt to adjust the coordination progression to the prevailing traffic speed. Since such speeds are highly variable, the resulting performance is often sub-optimal and network capacity is underutilized. Furthermore, the longer the road is, the smaller the bandwidth that can be achieved. Progressions typically break down after 10 or more signals, depending on distances and speeds. Researchers have proposed a variety of partition techniques in which an arterial is partitioned such that each sub-section is limited to a small number of signals [20][21][22][23][24]. These techniques attempt to determine the section lengths such that the overall sum of bandwidths is maximized. This breaks down the arterial progression into manageable units, yet vehicles are still stopped at the transition points and overall performance is not optimal. Yan et al. [25] propose a decomposition scheme for network-level green wave bands. Since the resulting bands may not be continuous for some arterials, they attempt to optimize connectivity as well. Yang and Ding [26] attempt to provide green wave control in a network using actuated logic.

The introduction of connected and automated vehicle (CAV) technologies offers new opportunities for traffic signal coordination. A number of survey papers present the issues that need to be addressed and possible solutions [27][28][24]. Tajalli and Hajbabaie [29] describe how to coordinate traffic signals with a mixture of connected automated vehicles and human-driven vehicles. This is an area of ongoing research and development.

In this paper a methodology is adopted that starts from a set of postulates modelled after Euclid. Green-arrow laws of motion are defined that allow the conversion of selected existing arterial roads into a network of roads that have the following properties: 1) vehicle platoons which travel with green-arrows make all traffic signals in a network of intersecting roads regardless of road length, and 2) have maximum bandwidth. Such a network is called a *Ride-the-Green-Wave* (RGW) road network. The mathematical model developed here applies to existing road networks that are topologically equivalent to a Cartesian Road network composed of either two-way roads, or alternate one-way roads, or a combination of the two road types. This methodology enables one to simultaneously: 1) get arbitrarily long progressions on existing suburban and urban road/street networks, and 2) obtain maximum throughput on these networks.

Section 2 introduces the basic concepts of the Euclidian approach: green-arrow laws of motion, Blocks, RGW-roads, proper green-arrow lengths, and describes the road-to-traveler-feedback-device. Section 3 shows that RGW-roads on existing two-way or alternate one-way roads that are topologically equivalent to a Cartesian Road network are possible. Section 4 shows that green-arrow laws of motion imply that the longest green-arrow on a network of two-way roads (alternate one-way roads) is one (two) Block(s). Section 5 shows that green-arrow laws of motion cannot be satisfied for some green-arrow lengths. Several green-arrow lengths that satisfy green-arrow laws of motion on two-way and alternate one-way road networks are identified in Section 6. Section 7 discusses the implications of green-arrow laws of motion for green-arrows that have unequal length on intersecting RGW-roads and their relationship to equal length green-arrows. Programs that were used to systematically find all proper green-arrow lengths on two-way or alternate one-way roads are described in Section 8 and Appendix B. The programs are used to find and tabulate all useful green-arrow lengths and to guess equations that provide all proper green-arrow lengths. Applications of green-wave theory start in Section 9 where it is shown how left-turn and reduced-flow-arrows can be used to accommodate anisotropic flow demand. Applications of the material presented here, a tabulation of new results, discussion and conclusions are given in Section 10.

## 2. Background

This section summarizes concepts that are needed for understanding the material presented here: how to get arbitrary long progressions with maximum bandwidth/throughput and uninterrupted flow on a network of two-way arterial roads. These concepts, briefly defined in this section, have been treated more fully [1]. Included are: the effect of green-arrows on traffic signals, green-arrow laws of motion, RGW-roads, road-to traveler-feedback-device (RTFD), left-turn-arounds, real RGW-nodes, blocks, Blocks, left-turn-arrows, reduced-capacity arrows, how concepts developed for a Cartesian road grid can be applied to existing road networks, virtual RGW-nodes, RGW-nodes, orphan-traffic-signals, singular points, and the art and science of creating RGW-roads.

*2.1 Effect of green-arrows on traffic signals*. The default state for all traffic signals is red. It is the motion of green-arrows which changes traffic signals to green, yellow, all-red and red. When a green-arrow leaves an intersection the traffic signal returns to its default state. The moment a green-arrow enters an intersection, it turns the traffic signal green in the green-arrow's direction of motion. As the green-arrow leaves an intersection it first turns the traffic signal yellow and then red. One can think of green-arrows as being graphical analogs of physical arrows with yellow tail feathers and a red colored shaft aft the tail feathers. When the yellow (red) portion of the arrow enters an intersection, it turns the traffic signal yellow (red). Green-arrows correspond to potential vehicle platoons.

*2.2 Green-arrow laws of motion*. The following four green-arrow laws of motion, the analogue of Euclid's postulates, are stated. 1) Green-arrows never stop moving, 2) Green-arrows fully utilize each intersection. 3) Green-arrows never intersect one another. 4) On a Cartesian-grid green-arrows move with a constant speed. Property one is introduced because we seek to coordinate traffic signals so that vehicle platoons never stop moving. Property 2 is needed because we seek maximum throughput/bandwidth. Property 3 is required because intersecting green-arrows corresponds to platoons of vehicles crashing into one another. Responsibility for the avoidance of vehicles crashing into one another, usually delegated to traffic signals, is here delegated to green-arrows.

Properties 1- 3 are essential properties of green-arrows. Green-wave theory explores mathematical consequences of these postulates. Property four applies only to a Cartesian Road grid and will be generalized for existing road networks in Section 2.8. However, property four is a natural postulate: 1) on a Cartesian-grid symmetry demands that speed in the north/south directions be the same as that in east/west directions and 2) regardless of where one is in a Cartesian grid since each square looks the same. This implies that unless vehicle speed is to vary depending on how far a vehicle is from an intersection, vehicle speed must be constant. Postulate four is useful: it enables uninterrupted maximum flow on Cartesian Road networks and is easily modified (Section 2.8) to get uninterrupted maximum flow on regular (Section 2.8) existing suburban roads [1].

*2.3 RGW-roads*. On RGW (Ride-the-Green-Wave) roads, vehicles in a green-wave (represented by a green-arrow) traveling at the recommended speed make every traffic signal regardless of how far they travel or their travel direction. RGW-roads are defined by two properties: 1) uninterruptible flow and 2) maximum throughput/bandwidth, i.e. each intersection is capable of being fully utilized in four directions. Alternately, RGW-roads are defined as roads where green-arrow laws of motion are satisfied.

Maximum flow in four orthogonal directions implies there is no time in the traffic signal cycle for direct left turns. Cloverleaf left-turns (Section 2.5) allow a left-turn capability and maximum flow in four orthogonal directions. When traffic signals are evenly spaced, as they are for a Cartesian Road grid, vehicles travelling at a constant speed make every traffic light. In the more usual case where traffic signals are not evenly spaced, vehicles vary their speed to make every traffic signal and are guided on how to do this by an RTFD.

*2.4 Road-to-Traveler-Feedback-Device (RTFD)*. Figure 1 illustrates how the RTFD, a smart phone application, communicates with the user assuming the user is a motorist. A paper currently in development will describe how and why the RTFD communicates with pedestrians and cyclists. The RTFD could also communicate with self-driving vehicles but how it would do this is outside the scope of this paper.

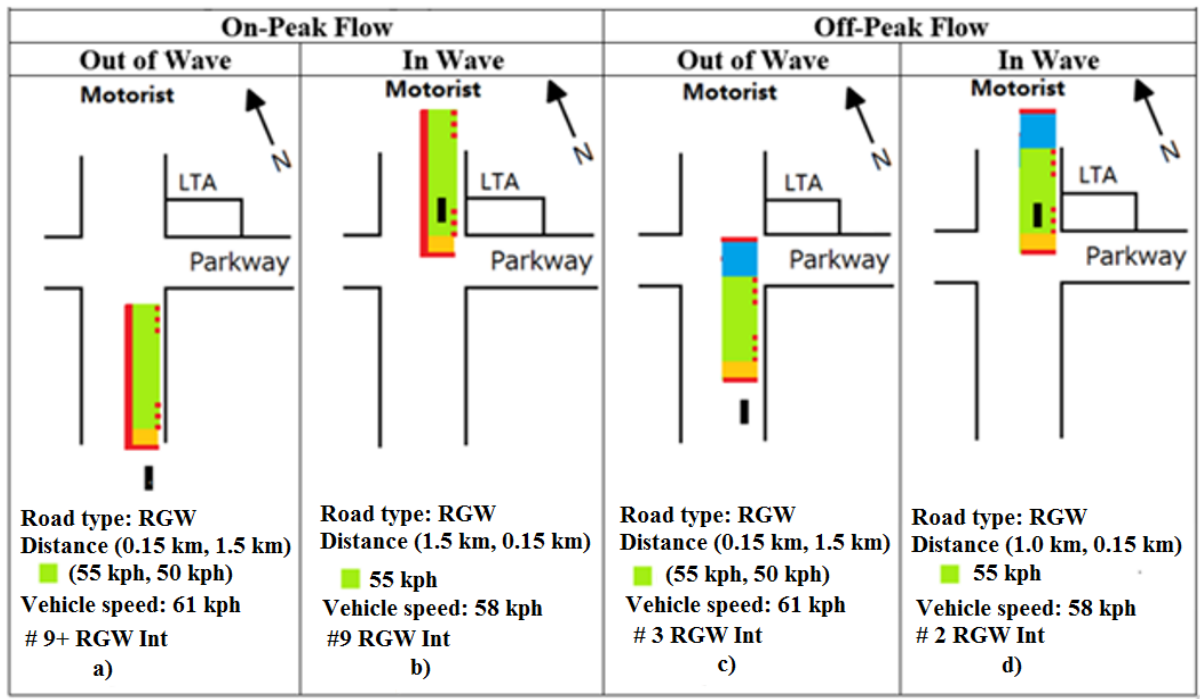

FIGURE 1. Representative RTFD motorist displays. a) motorist out of wave during on-peak flow. b) motorist in wave during on-peak flow. c) motorist out of wave during off-peak flow. d) motorist in wave during off-peak flow.

The display for peak vehicle flow is different than that for off-peak vehicle flow. The reason for this is that on two-way roads potential platoons of vehicles are potentially always going through a signalized intersection which implies there is no opportunity for motorists who want to make a direct left turn to do so. The solid vertical red line in Figure 1 a) and b) communicates this to drivers. The lack of a solid red line in Figure 1c and 1d indicates direct left turns can be made and that this can be done without stopping for vehicles in the blue portion of the c) and d) display. In Figures 1 a) and 1 c) the motorist, represented by a small black rectangle, is outside of the green-arrow and in Figures 1 b) and 1 d) the motorist is inside the green-arrow, here represented by a green rectangle. When red dots enter an intersection that means pedestrians have a walk signal and have the right-of-way over vehicles that want to make right or left turns. Motorists wishing to make a right turn prefer to be in a portion of a green-arrow without red dots. At the bottom of the display in Figures 1 a) and c) distance entries correspond to the green-wave ahead of and behind the motorist while distance entries in Figures 1 b) and d) indicate distance to the leading and trailing edge of the green-wave. Information in Figures 1 a) and c) enable motorists to make an informed decision regarding speeding up to get in the green-arrow ahead of the driver or slowing down to get in the green-arrow behind the driver. Information given in Figures 1b) and d) provide information that guide motorists so they can position themselves in a desired green-arrow position. Information at the bottom of the display informs the driver how many RGW-intersections the current green-arrow configuration will persist. This information is useful for motorists who wishes to position their vehicle at a particular position in a green-, left-turn-, or reduced-capacity-arrow (left-turn and reduced-capacity arrows are described briefly in Section 2.7 and more fully in Section 9).

*2.5 Left-turn-around*. In Figure 1, the left-turn-around is denoted with the letters LTA. Because RGW-roads are uninterruptable maximum throughput/bandwidth roads, during periods of peak vehicle flow there is no time in the traffic cycle for vehicles to make conventional left turns, here called direct left turns. During periods of peak vehicle flow left-turn-arounds are used to make left turns. Figure 2 illustrates what is meant by a left-turn-around and what is meant by a direct left turn and compares left-turn-arounds with a highway cloverleaf.

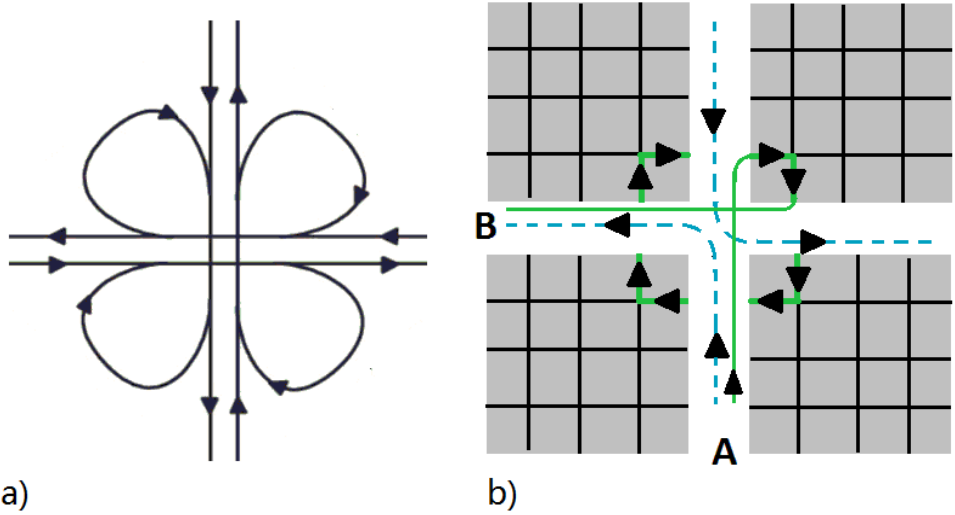

a)  b)

FIGURE 2. Direct and cloverleaf-left-turns. a) Highway cloverleaf. b) Cloverleaf for RGW-road.

In Figure 2b), A and B denote intersecting RGW-roads; solid black lines denote streets. Dashed blue lines indicate how northbound and southbound vehicles make direct left turns. The solid green line illustrates how a northbound vehicle makes a cloverleaf left turn. During periods of peak flow on two-way RGW-roads vehicles utilize each intersection in the forward direction completely, necessitating the need for cloverleaf left turns. Solid black lines in Figure 2 b) correspond to existing streets. The small gray squares enclosed by the streets are termed *blocks*. The large gray squares enclosed by RGW-roads are termed *Blocks*.

**2.6 Real RGW-Nodes, Blocks and green-wave speed.** Real RGW-nodes are the locations where RGW-roads intersect one another. For the case where green-arrows in four orthogonal directions are of equal length, green-arrows move with a speed $v_g$ given by

$$v_g = \frac{D}{T_g} \qquad (1)$$

where $D$ is the constant distance between adjacent nodes and $T_g$ is the duration of the green, yellow and all-red signals. This result is evident in Figure 3: green-arrows move a distance $D$ in a time $T_g$. For the Cartesian Road grid (Figure 3) $D$ is constant but for a non-Cartesian Road (Figure 5) this is not true.

Figure 3 or video-1 and Section 2.8 show that green-arrow laws of motion can be satisfied on Cartesian road networks and existing road networks that are topologically equivalent to a Cartesian road network. For simplicity, the yellow tail feathers and red portion after the tail feathers are omitted from Figure 3 and video-1.

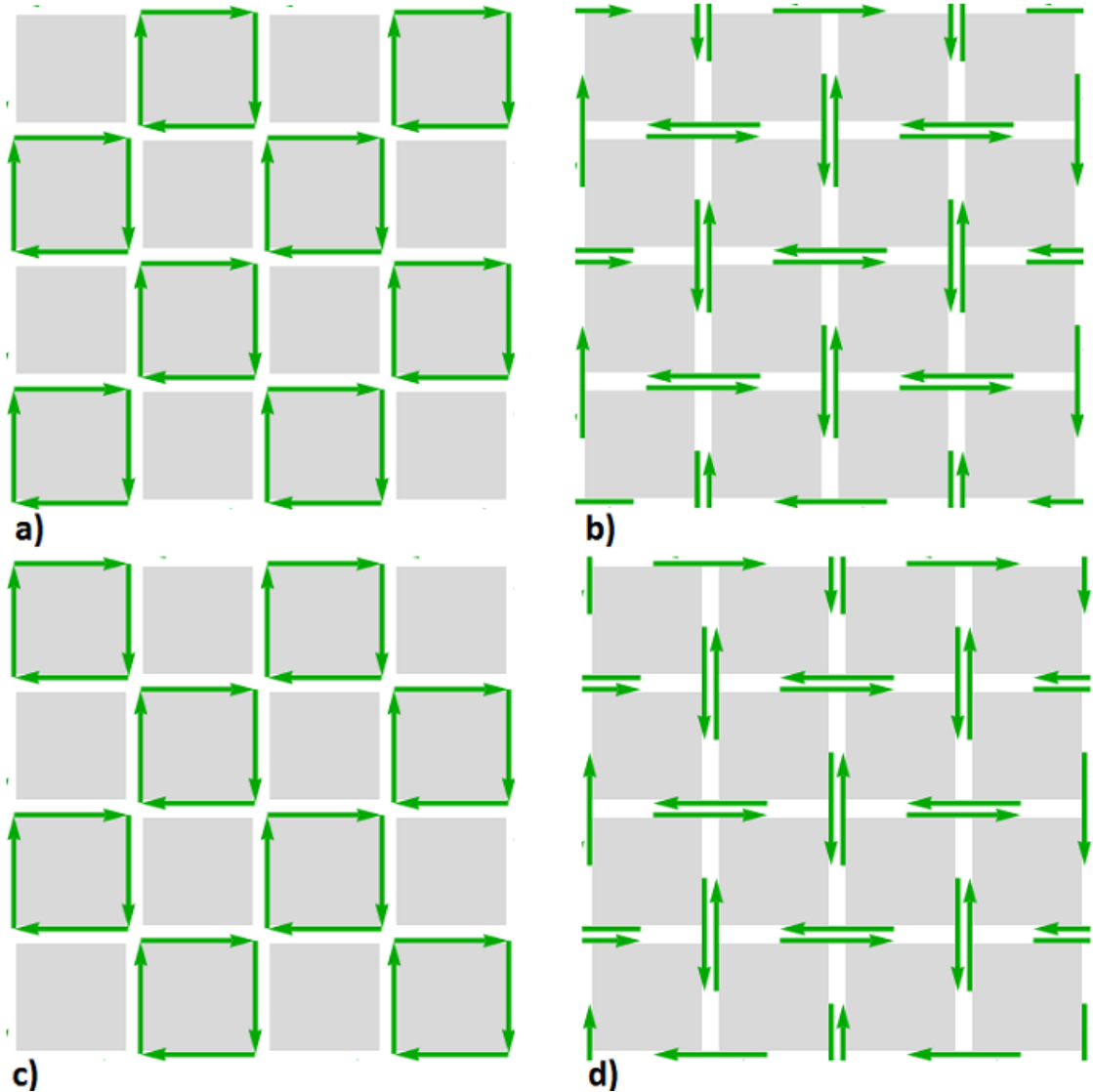

FIGURE 3. Illustration of green-wave/green-arrow motion on a Cartesian RGW-road network. a) Initial position of green-arrows. b) Green-arrows have advanced half a block from position a). c) Green-arrows have advanced an entire block from position a). d) Green-arrows have advance 1.5 blocks from position a). Advancement of green-arrows one half block from position d) reproduces a).

***Blocks*** are defined as the smallest areas enclosed by a network of RGW-roads. They are the gray areas illustrated in Figure 3 and Figure 2 b). This is different from the traditional definition of ***block*** as the smallest areas enclosed by a network of streets. Here, capital B is used to denote areas enclosed by RGW-roads and lower-case b is used to denote areas enclosed by a network of streets. As shown in Figure 2 b) typically a Block incorporates several blocks.

*2.7 Left-turn- and reduced-capacity-arrows*. A left-turn-arrow and a reduced-capacity-arrow are illustrated in Figure 4.

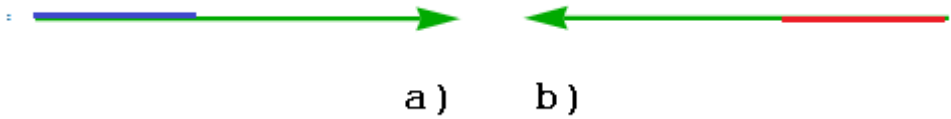

FIGURE 4. Left-turn-arrow and reduced-capacity-arrow. a) Left-turn-arrow. b) Reduced-capacity-arrow.

Left-turn- and reduced-capacity-arrows move like green-arrows, i.e. they obey green-arrow laws of motion (Section 2.2) but, as explained in Section 9.2, they have a different effect on traffic signals. When the blue portion of a left-turn-arrow enters an RGW-intersection it turns on a left turn signal for vehicles in the blue portion of the left turn arrow. Figure 4 illustrates a left-turn-arrow with the blue portion at the end of the arrow. Left-turn-arrows with the blue portion near the tip of the arrow are illustrated in Figure 1 c) and d). Vehicles in the green portion of a left-turn- or a reduced-capacity arrow make every traffic signal or can slow down to make a right turn. Vehicles in the blue portion of a left-turn arrow can make left turns at RGW-intersections without stopping. However, vehicles in the red-portion of a reduced-capacity arrow are stopped by the first traffic signal located at an RGW-intersection that the vehicle encounters. Left-turn- and reduced-capacity-arrows allow traffic engineers to adjust traffic signal timing to accommodate anisotropic flow demand [1].

*2.8 Method for applying Cartesian Road network results to existing road networks*. The Cartesian Road grid illustrated in Figure 3 is an abstraction used to coordinate traffic signals on a real road network illustrated in Figure 5 b)

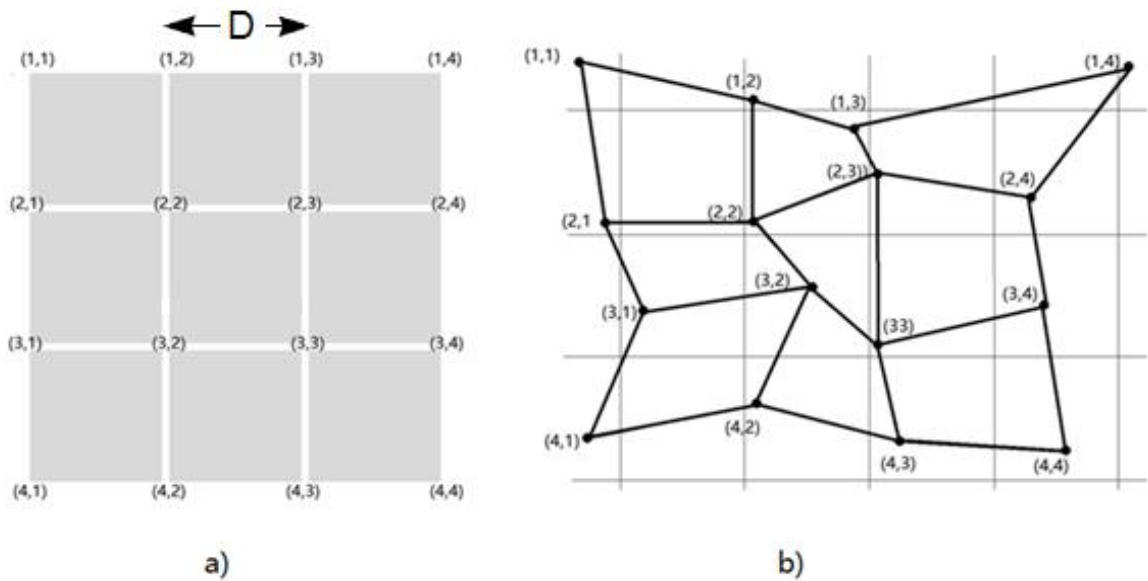

FIGURE 5. Generalizing RGW Cartesian Road grid traffic signal timing results to topologically-equivalent networks. a) Cartesian road network. b) Generalization of Cartesian Road network.

Figure 5 illustrates how Cartesian Road traffic signal timing results shown in Figure 3 and video-1 can be applied to a representative existing road network. Traffic signals at arbitrary nodes $(i, j)$ in Figure 5 b have the same state in a) and b), i.e. traffic signals at nodes (2,3) of the Cartesian grid and the generalized Cartesian grid are green, yellow and red at identical times. However, while the distance between adjacent nodes $D = D_0$ is constant in a) it is a variable in b) and this requires a different interpretation of the fourth green-arrow law of motion. Vehicle speed in the new fourth law of motion is given by (1) with the understanding that $D$ is the variable distance between adjacent nodes. To summarize: 1) traffic signals in a) and b) have the same state at all time, 2) vehicles in a) move at a constant speed, while vehicles in b) move at a variable speed given by (1). When conditions 1) and 2) are satisfied, vehicles in b) make every traffic signal. Figure 5 enables the discussion of existing road networks (exhibited in Figure 5 b)) using a Cartesian Road network (exhibited in Figure 5 a)).

*Plain or regular road network*. A road network is said to be *plain* or *regular* if it is topologically equivalent to a Cartesian Road network. Figure 5 b) is a plain/regular road network since it is topologically equivalent to the Cartesian Road network Figure 5 a).

*Cartesian road network and existing road networks*. It is apparent from Figure 5 that green-arrow properties valid for a Cartesian Road network are also valid for existing road networks that are plain. In the remainder of the paper results will be shown to be valid for Cartesian networks with the understanding that they are also valid for existing plain networks.

*Green-arrow lengths*. The lengths of green-arrows are always given in units of Blocks which makes green-arrow lengths dimensionless.

*Proper and improper green-arrow lengths*. Subsequently it will be found that on a Cartesian Road network with green-arrow length in the north-south direction equal to green-arrow lengths in the east-west direction, only certain green-arrow lengths can satisfy green-arrow laws of motion. Green-arrow lengths that satisfy green-arrow laws of motion are termed *proper* or *allowed* green-arrow lengths. Conversely, green-arrow lengths which cannot satisfy green-arrow laws of motion are termed *improper* or *not allowed* green-arrow lengths.

**2.9 Virtual RGW-nodes**. A problem with the fourth law of green-arrow motion (Section 2.6) is that when the length $D$ is large (for example between nodes (1,3) and (1,4) in Figure 5 b)) green-arrow speed $v_g$ can be excessive. Another problem with the fourth law of green-arrow motion is that when nodes are close together green-arrow speed can be excessively slow. This problem is solved by introducing virtual or imaginary RGW-nodes [1]. Virtual RGW-nodes are the locations of imaginary traffic signals which are introduced to slow down (speed up) green-arrow speed when it is excessive (to slow). Virtual nodes are also used to convert parity

-1 intersections to parity +1intersections (Section 2.21). Equation (1) applies to real and virtual nodes.

Virtual RGW-nodes have three functions: 1) they allow green-wave speed $v_g$ to have a reasonable value when the separation between nodes is excessive, 2) they allow green-wave speed to have a reasonable value when nodes are close together and 3) on an RGW-road network virtual RGW-nodes are needed to resolve traffic signal conflicts between north-south and east-west RGW-roads (Sections 2.21 and 2.22). The addition or removal of a virtual RGW-node will change a *negative parity* intersection to a *positive parity* intersection.

*2.10 Singular point*. In Figure 3 or [video-1](#) let $P'$ denote the point midway between nodes, i.e. the Block midpoint. Observe from Figure 3 or [video-1](#) that a green-arrow is always moving past point $P'$. This implies a traffic signal placed at point $P'$ would always be green in the forward direction which implies one cannot have either a cross-road or a cross-walk at point $P'$. Point $P'$ is called a singular point. Traffic signals cannot be placed at a singular point because they would never change state.

*2.11 Art and science of designing single RGW-roads*. The problem in designing RGW-roads is to appropriately choose the location of RGW-nodes and virtual RGW-nodes to get uninterrupted flow that approximates the speed limit while providing adequate green time for vehicles crossing the RGW-road. Once this has been done the program RGW-TLP calculates traffic signal time duration and offsets. The technique for appropriately choosing nodes and virtual nodes has been illustrated for a representative arterial road in Virginia, USA [1].

*2.12 Isotropic flow RGW-road* means that a single RGW-road has the ability to have potential maximal flow that is equal in opposite directions. Here isotropic flow is typically achieved using green-arrows that have the same length in opposite directions. An *isotropic flow RGW-road network* is a network of intersecting RGW-roads with the ability to potentially carry equal maximal flow in four directions. Isotropic flow on a network is typically achieved using green-arrows with the same length in four directions.

*2.13 Anisotropic flow RGW-road* means that a single RGW-road has a greater potential maximal flow in one direction than in the opposite direction. An *anisotropic flow RGW-road network* is a network of intersecting RGW-roads where the potential maximal flow is not equal in four directions.

*2.14 RGW-traffic and Orphan-traffic-signals*. RGW-traffic-signals are defined as traffic signals placed at the intersection of two RGW-roads. Orphan-traffic-signals are defined as traffic signals placed on an RGW-road at the location where a non-RGW-road or street intersects the RGW-road. An important property of orphan-traffic-signals is that their time offsets and duration are described by different equations than those of real or virtual RGW-traffic signals.

*Offset-time* $t_{offset}$ is the time it takes for a green-arrow to go from a reference node to an arbitrary real or virtual RGW-node. *Reduced-offset-time* $t_{roffset}$ for isotropic flow is defined by

$$t_{roffset} = Mod[t_{offset}, t_{cycletime}] \tag{2}$$

where $t_{cycletime}$ is the cycle time. For equal length green-arrows, green time in the forward direction $t_{gf}$ is given by

$$t_{gf} = t_{cycletime}/2 \tag{3}$$

Green time in the forward direction includes time the traffic signal is green, yellow and all-red. Equations (2) and (3) apply to traffic signals at nodes (real or virtual). The state of orphan-traffic signals is determined by different equations that depend on parameters $\xi$ and $d$. As indicated in Figure 6, the parameter $\xi$ is a dimensionless measure of how far away $x$ an orphan-traffic-signal is away from the nearest real or virtual node. The definition of $\xi$ implies $0 \leq \xi \leq 1/2$.

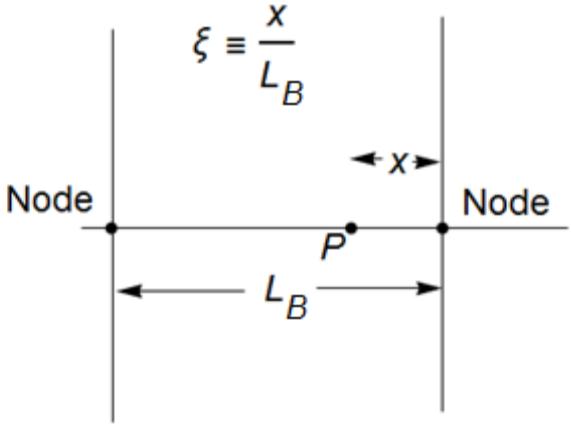

FIGURE 6. Definition of dimensionless parameter $\xi$.

In the definition of $\xi$ the approximation has been made that the distance $D$ between nodes is approximately $L_B$ the length of a Block. This approximation is valid when the separation between nodes is much greater than the width of the RGW-roads. Alternately, Block length can be defined as the distance between nodes (real or virtual).

The definition of the $d$ parameter requires knowing what is meant by a zero-offset node. At an instant of time on an RGW-road and at an RGW-intersection a *zero-offset node* is one where green-arrows in opposite directions are about to enter the intersection. Points O in Figure 7 illustrate what is meant by zero offset nodes on east-west RGW-roads. The same definition applies to north-south RGW-roads. Points A and B in the figure represent orphan-traffic-signals. The dimensionless parameter $d$ is defined as the distance of an orphan-traffic-signal from the nearest zero-offset node divided by Block length $D$. The definition of $d$, implies $0 < d < 1$. The red lines in Figure 7 correspond to singular points (mid-Block locations where traffic signals cannot be placed). Orphan-traffic-signals A satisfy the relationship $0 < d < 1/2$ while orphan-traffic-signals B satisfy the relationship $1/2 < d < 1$.

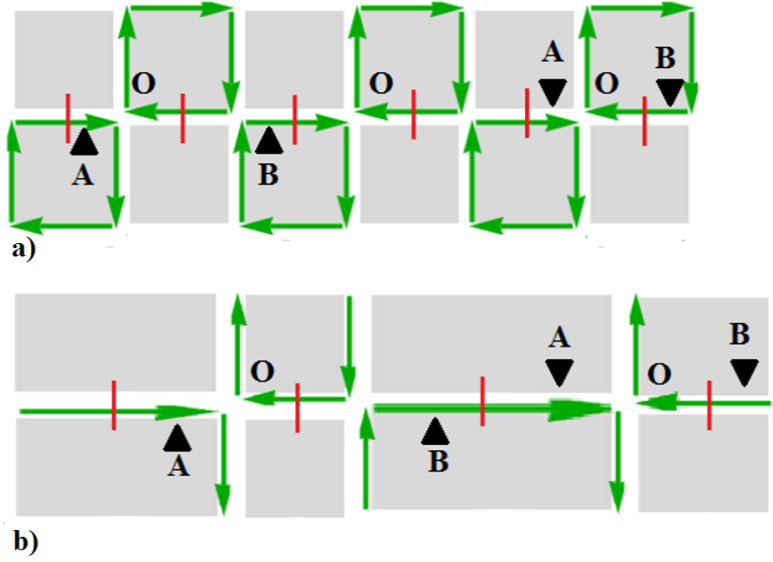

Figure 7. Illustration of *RGW-zero-offset-nodes* and *d parameter* for uniform and non-uniform Cartesian grids. Points A, B represent orphan-traffic signal locations on an east-west running RGW-road for a uniform and non-uniform road network. Points O, located where green-arrows are both about to enter the RGW-intersection, are termed *zero-offset-nodes*. a) Cartesian road network, b) non-uniform Cartesian road network.

With this background, reduced-offset-time orphan-traffic-signal equations in the isotropic flow case is

$$t_{roffset} = \begin{cases} T_g(1-\xi) & 0 < d < 1/2 \\ T_g(2-\xi), & 1/2 < d < 1 \end{cases} \quad (4)$$

where $T_g$ is the time when a traffic signal is green, yellow and all-red at a real or virtual node. For the isotropic flow case $T_g$ equals half the cycle time. The parameter $t_{roffset}$ describes when in the cycle time orphan-traffic signals-turn green. The duration of the green traffic signal on RGW-roads is

$$T_{gf}(\xi) = (1 + 2\xi)T_g, \quad 0 \leq \xi < 1/2 \quad (5)$$

and the duration of the green traffic signal in the cross direction to the RGW-road is

$$T_{gx}(\xi) = (1 - 2\xi)T_g, \quad 0 \leq \xi < 1/2 \quad (6)$$

As a check, observe that $T_{gf} + T_{gx} = 2\,T_g = t_{cycletime}$, an obvious result for isotropic flow.

The method used to derive (5) and (6) follows. Result (6) was derived by assuming $T_{gx}(\xi)$ is a linear function of $\xi$. Then the conditions that $T_{gx}(1/2) = 0$ and $T_{gx}(0) = T_g$ determines $T_{gx}(\xi)$. The condition that $T_{gf} + T_{gx} = 2\,T_g$ determines $T_{gf}(\xi)$. The assumption that $T_{gx}(\xi)$ is a linear function of $\xi$ was checked by a small computer simulation and (4) - (6) were confirmed by RGW-SIM [1]. The derivation of (4), (5) and (6) is given in [1].

*2.15 Singular interval*. Subsequent analysis will show that the singular point observed for isotropic flow spreads out to an interval when green-arrows in orthogonal directions are of different length. Stated differently, on the RGW-road with the longer green-arrows, instead of their being a single point where a traffic signal in the cross direction never turns green, there is an interval on the RGW-road where a traffic signal in the cross direction never turns green.

*2.16 Forbidden interval*. The forbidden interval is an interval that includes a singular point or a singular interval where it is impractical to place a traffic signal because green time in the cross direction is too small for vehicles or pedestrians to cross the RGW-road. The difference between a singular interval and a forbidden interval is that a traffic signal placed in a singular interval never turns green in the cross direction whereas in a forbidden interval the traffic signal in the cross direction turns green for a period too small to be useful.

*2.17 Green-arrow spatial periodicity*. Examination of Figure 3 a) shows that the initial placement of a green-arrow results in a green-arrow pattern that is spatially periodic with a period of two blocks and that this spatial periodicity is maintained for all time. On a Cartesian one- or two-way RGW-road network green-arrows are always spatially periodic regardless of whether green-arrows in orthogonal directions are of equal or unequal length.

*2.18 Green-arrows are restricted to discrete lengths*. Examination of Figure 3 shows that green-arrows have a length of 1. Recall that green-arrow lengths are dimensionless since green-arrow length is defined as the length of the green-arrow divided by the length of the Block it is in. Subsequent analysis shows that when green-arrows in orthogonal directions are of equal length, then green-arrow lengths are restricted to specific values (sections 4, 5 and 6). When green-arrows in orthogonal directions are unequal in length, then green-arrow laws of motion restrict the sum of the green-arrow lengths on intersecting RGW-roads to discrete values (Section 7).

*2.19 Maximum green-arrow length*. It is well known that two waves with the same amplitude traveling in opposite directions result in a standing wave. In a jump rope or a pipe organ, one standing wave has a lowest frequency (longest wavelength) called the fundamental and other frequencies which are multiples of the fundamental frequency are called the first, second, ..., $n^{th}$ harmonic. In the case of any particular two-way RGW-road, green-waves are sent in two opposite directions so it is not surprising that standing waves are observed and one of those standing waves has the longest wavelength. Subsequently it will be shown that on a Cartesian

Road grid the maximum length equal length green-arrow for alternate one-way arterial roads is two and the maximum equal length green-arrow length for two-way roads is one. Green-arrow lengths are often quoted as dimensionless numbers because they are always the length of the green-arrow divided by the length of the Block and both quantities have the unit of length.

*2.20 Importance of green-arrow length to traffic engineers*. The longest length green-arrows are especially interesting to traffic engineers because long length green-arrows imply the smallest possible fraction of the traffic cycle time is used on yellow and all-red lights. Green-arrows shorter than the fundamental are also of interest to traffic engineers since they allow for shorter red light wait times at the expense of less efficient traffic flow. Short green-arrow lengths are also useful for preventing left-turn waiting lane overflow for roads that do not have left-turn-arounds. The length of red light wait time on RGW-roads is of less importance than on ordinary arterial roads because vehicles traveling at the recommended speed on these roads do not have to stop for red lights.

Additional concepts not discussed in [1] are useful for obtaining maximum throughput and uninterrupted flow on existing two-way arterial road networks.

*2.21 Parity of intersections*. Intersections where green-arrows in orthogonal directions do not collide are said to have parity $+1$; otherwise, the intersection has parity $-1$. Examples of intersections with parity $+1$ and $-1$ are shown in Figure 8.

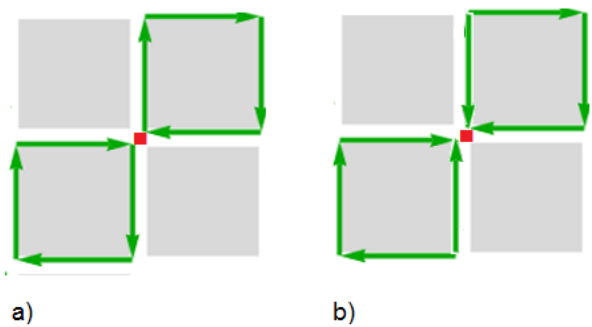

FIGURE 8. Parity $+1$ and parity $-1$ intersections. a) The intersection with the red square has parity $+1$. b) The intersection with the red square has parity $-1$.

*2.22 Technique for applying green-wave theory to existing road networks*. The technique for applying green-wave theory to an existing two-way arterial road has been described [1]. This technique is applied individually to each RGW-road in the road network using the method of [1]. At some places in the road network where RGW-roads intersect the parity is $+1$ and there is no need for further action. At other places where RGW-roads intersect the parity is $-1$. This is addressed by either adding or taking away a virtual node. Adding or taking away a virtual node changes a $-1$ parity node to a parity $+1$ node. The *reconciliation* technique for changing each $-1$ parity nodes to a parity 1 nodes in such a way that all nodes are parity 1 nodes is described in a paper being prepared for publication.

**2.23 Equal length green-arrows** are green-arrows that have the same length in orthogonal directions on a Cartesian Road network. When green-arrows on intersecting roads have the same length, we refer to this as an isotropic flow network.

**2.24 Unequal length green-arrows** are green-arrows that have different lengths in orthogonal directions on a Cartesian Road network. When green-arrows on intersecting roads have unequal length, we refer to this as an anisotropic flow network.

## 3. Green-arrow laws of motion can be satisfied.

Section 3.1 demonstrates: 1) that green-arrows with a length of *two* satisfy green-arrow laws of motion on an *alternate one-way* Cartesian Road network and 2) that green-arrow laws of motion imply that the placement of a single green-arrow on the road network at an arbitrary instant of time determines the placement of all green-arrows on the network regardless of network size for all time. Section 3.2 demonstrates: 1) that green-arrows with a length of *one Block* satisfy green-arrow laws of motion on a *two-way* Cartesian Road network and 2) that green-arrow laws of motion imply that the placement of a single green-arrow on the *two-way* road network at an arbitrary instant of time determines the placement of all green-arrows on the network for all time. Although the demonstrations are for a Cartesian Road network Section 2.8 implies these results are applicable to existing plain road networks.

*3.1 Demonstration of uninterrupted isotropic maximum throughput on arbitrarily long alternate one-way arterial roads*. Figure 9 or [video-2](video-2) demonstrate that green-arrow laws of motion can be satisfied on an arbitrarily large Cartesian network of alternate-one-way roads. Here green-arrows are of length two. In Section 4.1 it will be shown that green-arrow laws of motion on plain network of alternate one-way roads cannot be satisfied with green-arrow lengths longer than two.

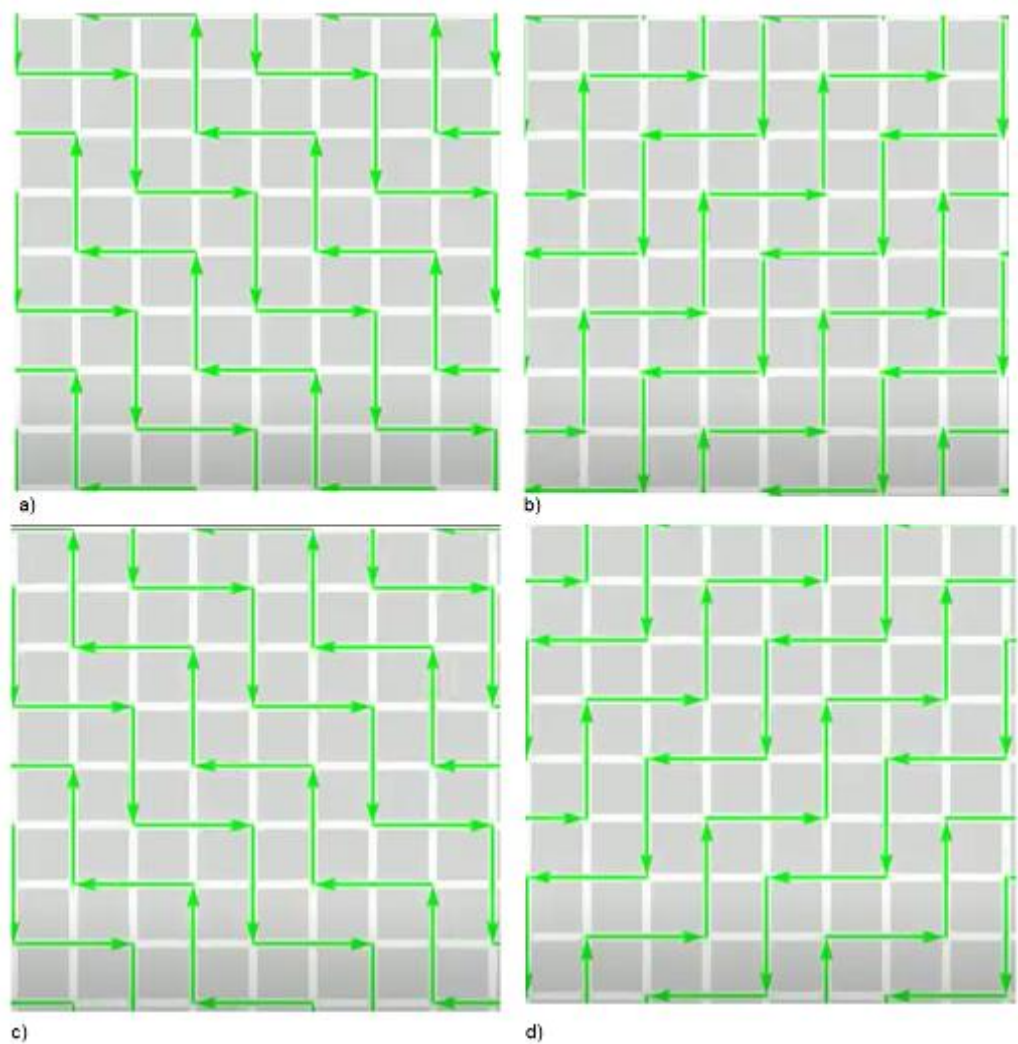

FIGURE 9. Top-down view of alternate one-way RGW-road network with green-arrows two Blocks long. a) Initial placement of green-arrows. b) Green-arrows have advanced one Block from a). c) Green-arrows have advanced two Blocks from a). d) Green-arrows have advanced three Blocks from a).

Observe that green-arrows moving from a) to b) to c) to d) fully utilize each intersection and did not stop moving and did not intersect one another, i.e. green-arrows in Figure 9 obey green-arrow laws of motion. The advancement of green-arrows in d) by one Block reproduces Figure 9 a) which implies traffic signals controlled by green-arrows are periodic in time. Also observe that the green-arrow pattern is periodic with a spatial period of four Blocks. The periodicity of green-arrows in space and time implies green-arrows can move forever over the entire plane and for all time fully utilizing each intersection fully without stopping and without colliding. Because the green-arrow pattern shown in Figure 9 a) can be arbitrarily large, the traffic light coordination method shown in this figure applies to plain one-way road networks in cities of arbitrary size.

Figure 10 demonstrates that green-arrow laws of motion imply that the placement of one RGW-arrow on an *alternate one-way* road network at one instant of time determines the placement of all green-arrows over the entire two-dimensional grid for all time. This important result is used repeatedly to derive results and to automate the process of finding proper green-arrow lengths.

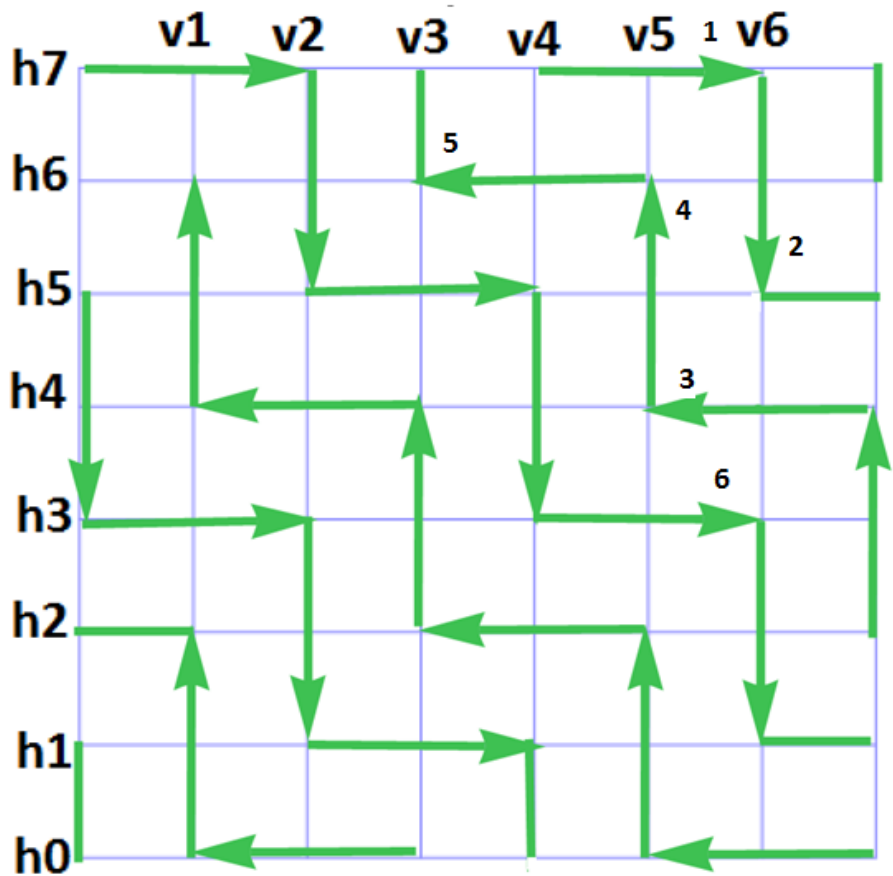

FIGURE 10. Illustrating that the placement of one green-arrow at one instant of time on a network of alternate one-way roads determines the location of all green-arrows over the entire plane for all subsequent time.

In Figure 10 the blue lines represent a top-down view of a system of alternate one-way RGW-roads. Start by placing green-arrow 1 entering intersection h7-v6 traveling in the eastern direction. The condition of maximum flow at intersection h7-v6 requires that a north- or south-bound green-arrow leave this intersection. The choice is determined by the v6 street direction which in this case is southbound. Hence the placement of green-arrow 2 entering intersection h5-v6. The maximum flow condition and the requirement that all waves travel at the same speed requires a western moving arrow 3 enter intersection h4-v5 to insure maximum flow at intersection h4-v6. This, in turn, requires a northern moving arrow 4 enter intersection h6-v5 which in turn requires a western moving arrow 5 enter intersection h6-v3. The requirement that intersection h3-v6 be fully utilized requires the placement of arrow 6. Clearly this process can be continued to produce the entirety of Figure 8 which represents a snapshot of green-arrow positions at an instant of time. The condition that green-arrows on a Cartesian Road network advance at constant speed determines the position of all green-arrows for all time. Given the street direction of the illustrated alternate one-way road network, the same pattern is reproduced if one starts with any arrow in Figure 8.

In Figure 10 the first green-arrow was placed with its tip at an intersection. This simplifies the placement of other green-arrows. Had the first green-arrow not been placed at an intersection the location of all other green-arrows would have still been determined but their graphical placement would have been more difficult.

### 3.2 *Demonstration of uninterrupted isotropic maximum throughput on two-way arterial roads*

Figure 11 or video-3 demonstrate that uninterrupted isotropic maximum throughput is possible on a Cartesian network of two-way roads regardless of its extent. Green-waves have been discussed [4] with waves moving at different speeds in the north/south and east/west directions.

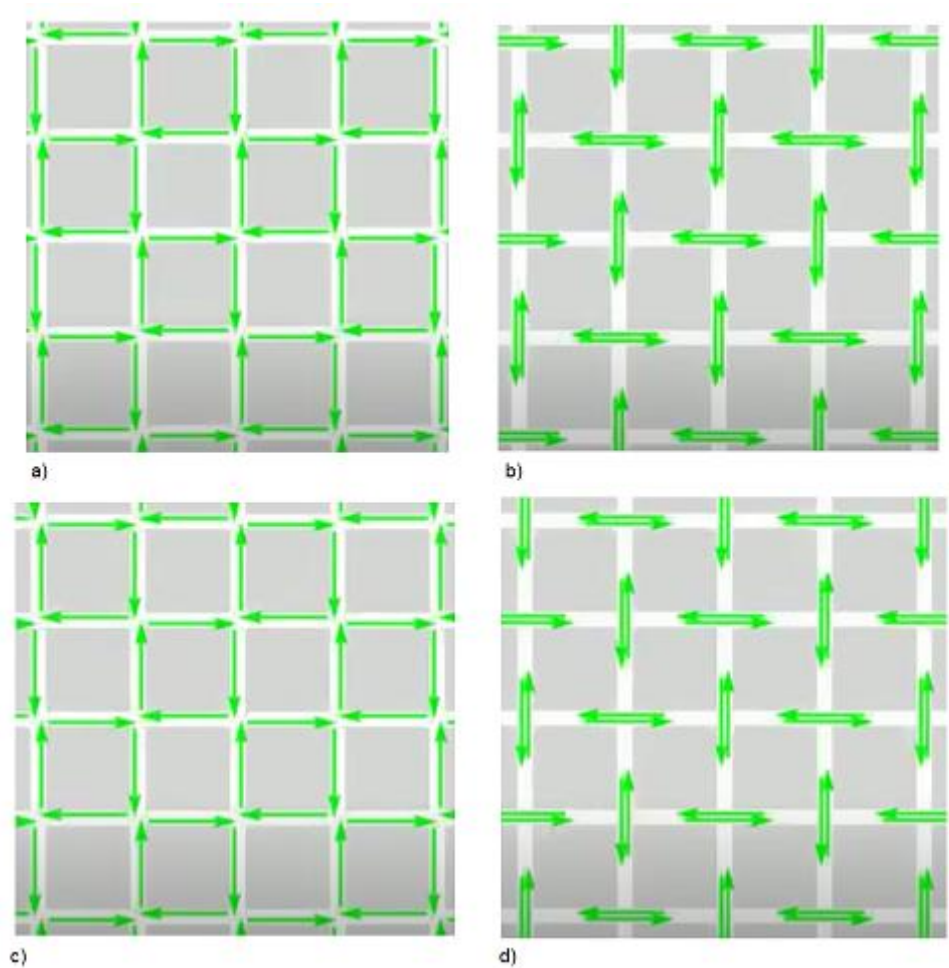

FIGURE 11. Top-down view of two-way RGW-road network with green-arrows 1 Block long. a) Initial placement of green-arrows. b) Green-arrows have advanced 1/2 Block from a). c)

Green-arrows have advanced one Block from a). d) Green-arrows have advanced 3/2 Blocks from a).

Observe that green-arrows moving from a) to b) to c) to d) in Figure 11 fully utilize each intersection and do not stop moving and do not intersect one another, i.e. they satisfy green-arrow laws of motion. The advancement of green-arrows in d) by 1/2 Block reproduces Figure 11 a) which implies traffic signals change periodically in time on two-way roads topologically equivalent to those illustrated in Figure 11. Also observe that the green-arrow pattern is spatially periodic with a period of two Blocks. This implies green-arrows can move forever over the entire plane and for all time fully utilizing each intersection without stopping and without colliding. This concludes the argument that two-way arbitrarily large uninterrupted flow maximum bandwidth/throughput plane road networks are possible with green-arrows of length one.

In Figure 11 or video-3 notice that a green-arrow is always moving at each mid-Block position. Thus, if one placed a traffic signal at a mid-Block position it would never turn green in the cross direction allowing pedestrians or cross traffic from an orphan traffic signal at the mid-Block position to cross the RGW-road. This is expressed by (6) in the limit where $\xi \to 1/2$. The mid-Block positions for each Block in Figures 7 and 11 are termed *singular points* because traffic signals at those locations never turn green in the cross direction.

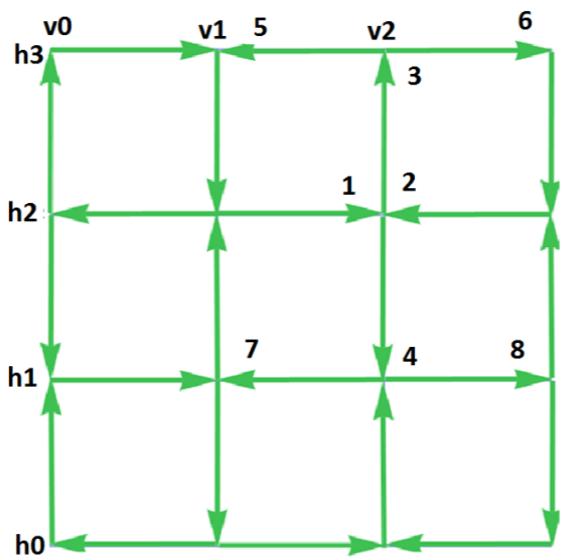

FIGURE 12. Illustration of result that the placement of one length one green-arrow on a Cartesian two-way road grid at one instant of time determines the subsequent placement of all green-arrows for all time.

Figure 12 shows that green-arrow laws of motion imply that the placement of a single green-arrow on a *two-way* road network at one instant of time determines the placement of all green-arrows on the network for all instants of time. Place the first green-arrow on road h2 with its tip at the $(h2, v2)$ intersection. Arrow 2 must be placed as shown to fully utilize the $(h2, v2)$ intersection. Green-arrows 3 and 4 must be placed as shown to fully utilize the $(h2, v2)$ intersection. Green-arrows 5 and 6 must be placed as shown to fully utilize the $(h3, v2)$ intersection and green-arrows 7 and 8 must be placed as shown to fully utilize the $(h1, v2)$ intersection. Continuing in this way the reader can easily confirm that each green-arrow must be placed as shown. The reader can also confirm that Figure 12 is reproduced regardless of what green-arrow in Figure 12 is placed first. Since the distance between adjacent nodes in Figure 12 is constant, (1) implies all green-arrows in this figure move with the same speed. Note that green-arrows are spatially periodic with a period of two Blocks. This implies the green-arrow pattern shown in Figure 12 can be extended to the entire plane. Note that green-arrow placement in Figure 12 is identical to green-arrow placement in Figure 11 a). This implies green-arrows in Figure 12 are temporally periodic with a period of $t_{cycletime}$.

## 4. Longest equal length green-arrows on two-way and alternate one-way RGW-roads.

The previous section demonstrated length two green-arrows satisfy green-arrow laws of motion (Figures 9 and 10) on an alternate *one-way* road network that is topologically equivalent to a Cartesian Road network. Length one green-arrows were shown to satisfy green-arrow laws of motion (Figures 11 and 12) on any existing *two-way* road network topologically equivalent to a Cartesian Road network. In Section 4 it is shown that green-arrows longer than two are not possible on a plain network of alternate one-way roads and that green-arrows longer than one are not possible on a plain network of two-way roads. The importance of these results: in Section 8 these results are used to find all proper green-arrow lengths.

*4.1 Alternate one-way RGW-roads.* Figure 13 shows a top-down view of an alternate one-way road network. If the blue lines are thought of as RGW-roads then the gray lines designate fractions of an RGW-road and can also be thought of as streets.

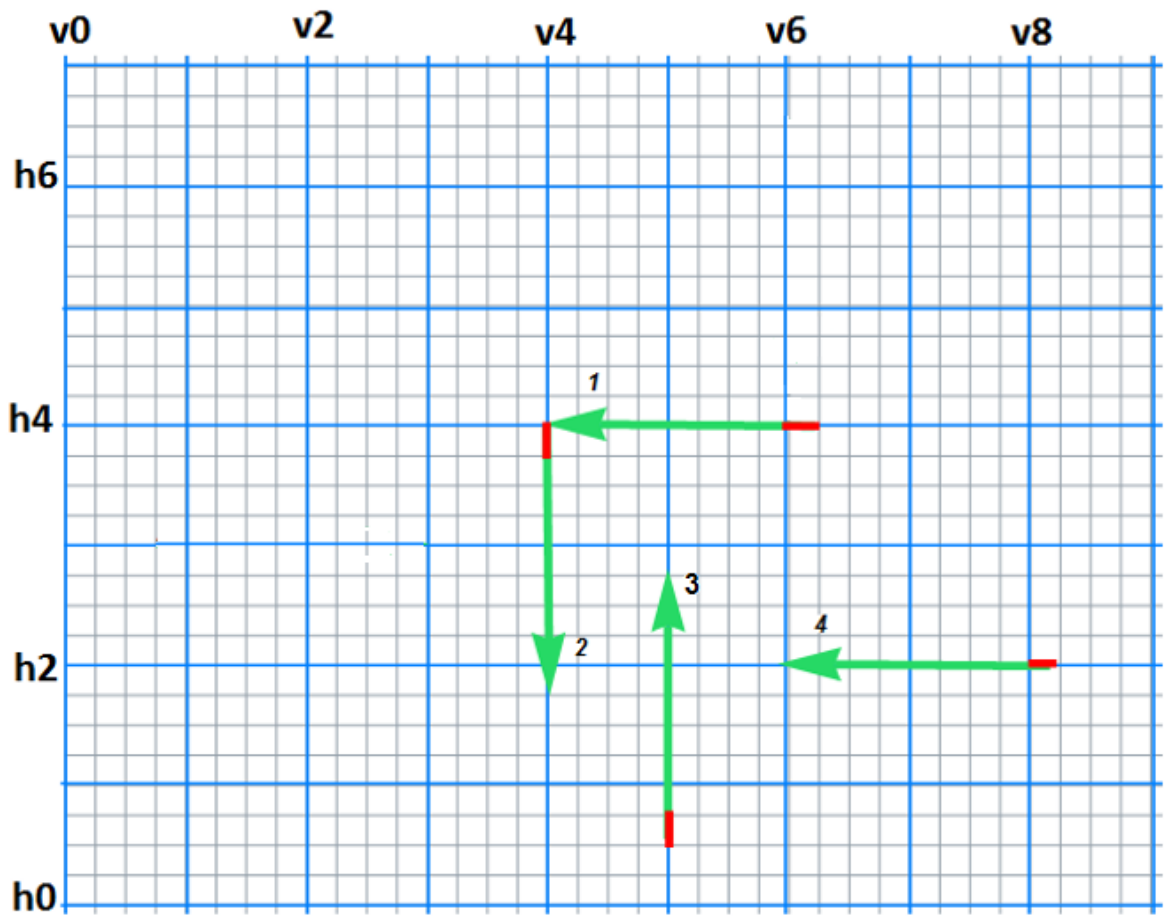

FIGURE 13. Equal length green-arrows longer than two cannot be placed on an alternate one-way Cartesian Road network.

It will now be shown that equal length green-arrows longer than two are not possible on a Cartesian grid of alternate one-way roads. To simplify the exposition this is shown for green-arrows 2 ¼ Blocks long but the argument works for any green-arrow length longer than 2 Blocks. Although the argument is given for a Cartesian Road network it is applicable to existing road networks that are plain. In Figure 13 the first green-arrow of length 2 ¼ Blocks is placed on RGW-road $h4$ with its head at the intersection $(h4, v4)$. Green-arrow 2 must be placed as shown to fully utilize intersection $(h4, v4)$ and green-arrow 3 must be placed as shown to fully utilize intersection $(h4, v5)$. Green-arrow 4 must be placed as shown to fully utilize intersection $(h2, v4)$. However green-arrows 3 and 4 collide at intersection $(h2, v5)$. This shows that green-arrows with length 2 ¼ cannot satisfy green-arrow laws of motion on an alternate one-way Cartesian Road grid. Although larger graphs would be needed, the same argument could be used with any green-arrow of length between 2 and 2 ¼. Also, the same argument works for green-arrows longer than 2 ¼. This concludes the demonstration that green-arrows longer than 2 are not possible on an alternate one-way Cartesian or plain road network.

*4.2 Two-way roads*.  In this section it is shown that green-arrows longer than 1 cannot satisfy green-arrow laws of motion on a Cartesian network of two-way roads.  Figure 14 shows a top-down view of a Cartesian RGW-network of two-way roads.  When the blue lines are interpreted as two-way arterial RGW-roads then gray lines are interpreted as streets (or as fractions of a Block).

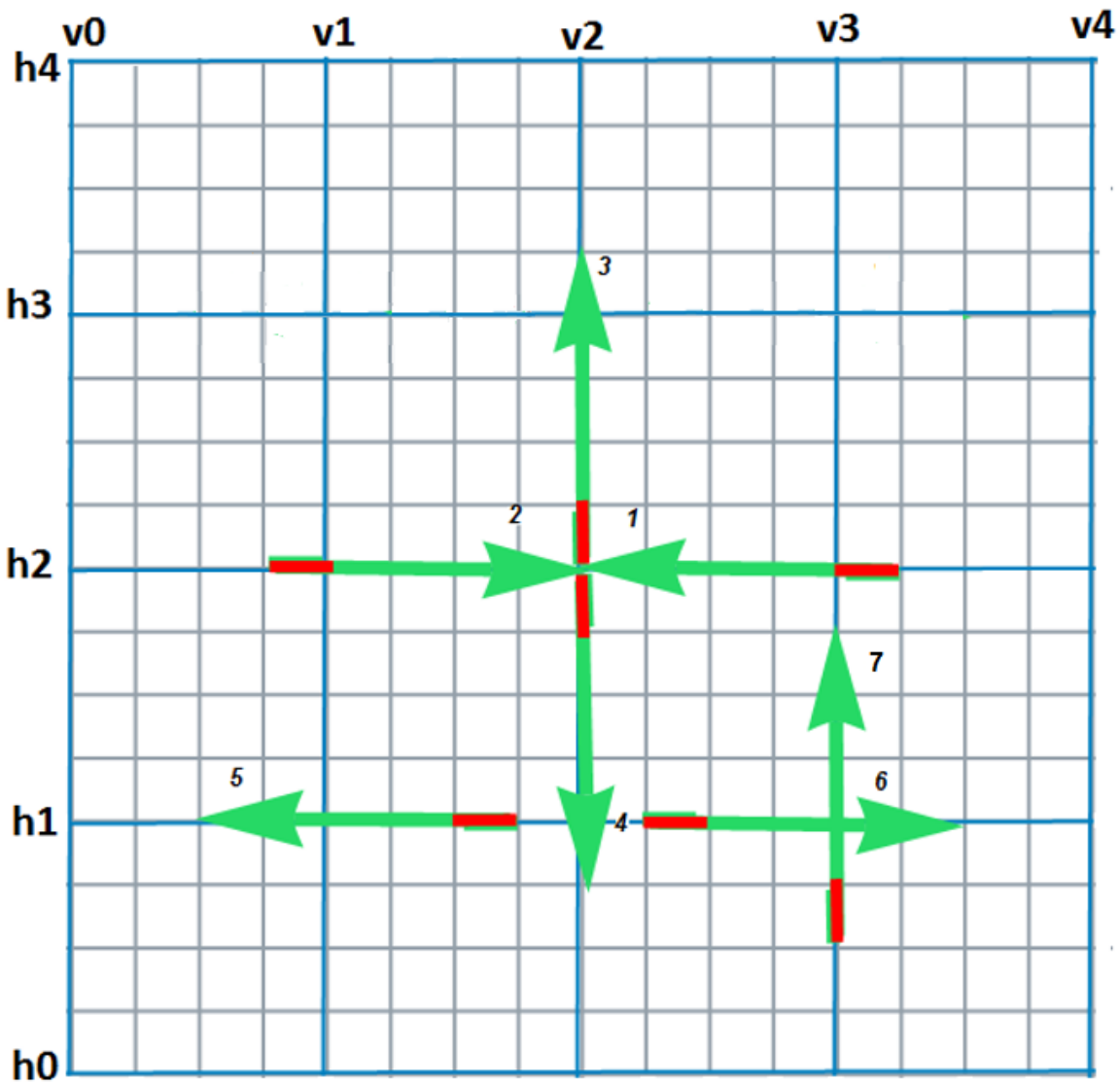

FIGURE 14.  Green-arrows longer than 1 are not possible on a Cartesian network of two-way equal green-arrow length RGW-roads $h0, \ldots, v4$.  Red lines indicate the excess $x$ over one for green-arrow length.  Here $x = ¼$ .

Using Figure 14 it will now be shown that on a two-way Cartesian Road network as long as $x > 0$ (x is green-arrow length greater than one) green-arrow laws of motion cannot be satisfied.  Green-arrow 1 is arbitrarily placed as shown.  The requirement that all intersections be fully utilized then immediately implies the placement of green-arrows 2, 3, and 4.  Green-arrows 5 and 6 must be placed as shown to fully utilized intersection$(h1, v2)$.  Green-arrow 7 must be placed as shown to fully utilize intersection ($h2, v3$) but this then causes green-arrows 6 and 7 to collide which violates the third green-arrow law of motion. Clearly green-arrows intersect for $0 < x \leq 1/4$. A similar argument shows green-arrows intersect when ¼< $x$ < 1. This demonstrates there are no green-arrows with length greater than 1 and less than 2 which satisfy green-arrow laws of motion.

We now show that no green-arrows with length greater than 2 can satisfy all green-arrow laws of motion on a two-way road network. In Section 4.1 it was demonstrated that green-arrows greater than 2 were not allowed for a Cartesian RGW-road network of alternate one-way roads. A Cartesian system of two-way RGW-roads is a generalization of a Cartesian system of one-way RGW-roads. Because green-arrows longer than 2 are not possible on a system of alternate one-way roads, they are not possible on a system of two-way roads. We conclude that the longest green-arrow length is 1 for isotropic flow on a Cartesian grid or plain network of two-way roads.

## 5. Forbidden equal length green-arrow lengths on RGW-roads.

Section 3 showed that equal length green-arrows of length 1 satisfy green-arrow laws of motion on a Cartesian road-network of two-way roads and that equal length green-arrows of length 2 satisfy green-arrow laws of motion on a Cartesian alternate one-way road network. Section 4 showed that equal length green-arrows longer than 1 cannot satisfy green-arrow laws of motion on a Cartesian road-network of two-way roads and that equal length green-arrows longer than 2 cannot satisfy green-arrow laws of motion on a Cartesian road-network of alternate one-way roads. Section 2.8 showed that results for Cartesian Road networks can be applied to topologically equivalent real road networks. This section demonstrates that some equal length green-arrow lengths cannot satisfy green-arrow laws of motion on a Cartesian grid or on existing plain road networks. These are termed *forbidden green-arrow lengths*. All proofs are done using Cartesian Road networks however, Figure 5 implies results shown here apply to existing plain road networks.

*5.1 Green-arrow length of 3/2 on alternate one-way roads*. In this section it is shown that a green-arrow length of 3/2 cannot satisfy green-arrow laws of motion on an alternate one-way Cartesian Road network. It is reasonable to think green-arrows of length 3/2 are possible on an alternate one-way Cartesian Road since it was shown in Section 3.1 that green-arrows of length 2 satisfy green-arrow laws of motion. The method of proof: 1) place one green-arrow down at one instant of time; 2) the condition that each intersection be fully utilized requires other green-arrows be placed in certain mandated positions at that instant of time; 3) when placed in the mandated positions, two green-arrows intersect, which demonstrates that this particular green-arrow length is not allowed.

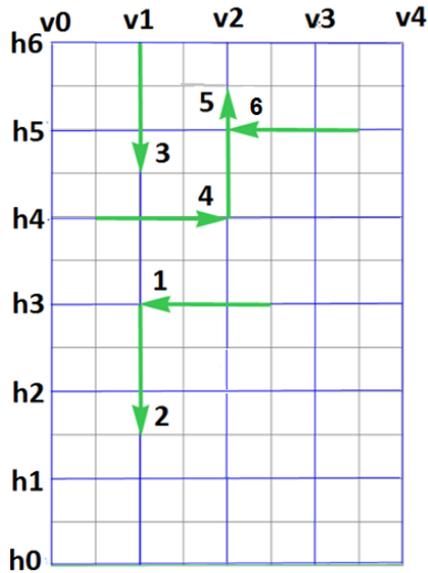

FIGURE 15. Demonstration that green-arrows of length 3/2 are not allowed on a Cartesian alternate one-way road network.

In Figure 15 place green-arrow 1 so that its head is entering intersection $(h3, v1)$. Then to maintain maximum flow at intersection $(h3, v1)$ green-arrows 2 and 3 must be placed as shown (assuming $v1$ is southbound). Green-arrow 4 must be placed as shown to fully utilize intersection $(h4, v1)$ and green-arrow 5 must be placed as shown to fully utilize intersection $(h4, v2)$. Green-arrow 6 must be placed with its head entering intersection $(h5, v2)$ to fully utilize intersection $(h5, v1)$. However, this results in an intersection of green-arrows 5 and 6. This shows that a green-arrow length of 3/2 is not allowed on a Cartesian grid of alternate one-way roads.

*5.2 Green-arrow of length 3/2 on two-way roads*. This subsection shows that equal length green-arrow lengths of 3/2 are forbidden on a Cartesian Road grid. The argument given in Section 4.2 showed equal length green-arrows on a Cartesian grid of two-way roads with length greater than 1 cannot satisfy green-arrow laws of motion so it might seem that this subsection is not necessary. This specific case is presented to show the result shown in Section 4.2 is valid for a specific case and to illustrate the method used in programs described in Section 8 and Appendix B.

In Figure 16 green-arrow 1 is placed so its head is at intersection $(h3, v1)$ at some instant of time. Green-arrows 2, 3 and 4 are placed as shown to fully utilize intersection $(h3, v1)$. Green-arrow 5 is placed to fully utilize intersection $(h1, v1)$. Green-arrows 5, 6 and 7 are placed to fully utilize intersections $(h1, v1)$ and $(h1, v2)$. Green-arrow 8 must be placed as shown to fully utilize intersection $(h2, v2)$. However, green-arrows 8 and 4 intersect each other. This demonstrates equal length green-arrows of length 3/2 cannot satisfy green-arrow laws of motion on a two-way Cartesian Road network.

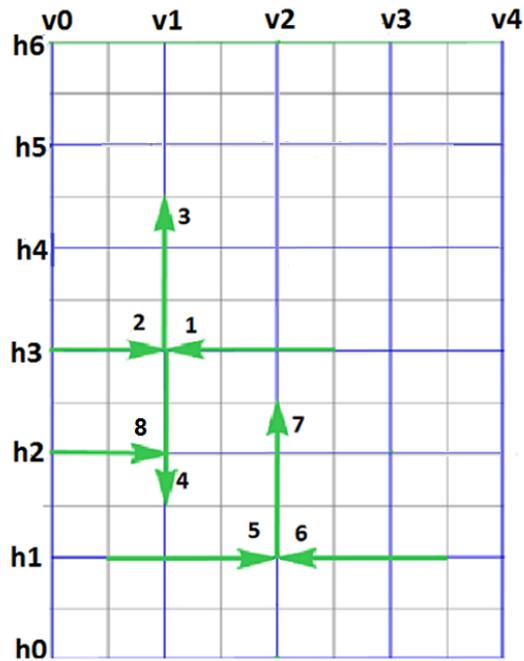

FIGURE 16. Demonstration that green-arrows of length 3/2 are not allowed on a Cartesian two-way road network.

## 6. Allowed equal length green-arrow lengths on two-way and alternate one-way road networks.

In Section 3 it was shown that on a Cartesian alternate one-way road network with equal length green-arrows, green-arrow laws of motion can be satisfied with green-arrows of length 2 on an alternate one-way system of roads and they could be satisfied with equal length green-arrows of length 1 on a Cartesian two-way road network. In Section 4 it was shown that equal length green-arrows greater than 2 are not possible on an alternate one-way system of roads and that equal length green-arrow with length greater than 1 are not possible on a two-way Cartesian Road network. Section 5 showed that with equal length green-arrows some green-arrow lengths less than the maximum value cannot satisfy green-arrow laws of motion on either two-way or alternate one-way road networks. Section 2.8 showed that these results are also true for road networks topologically equivalent to a Cartesian Road network. This section uses the method illustrated in Figures 10 and 12 to find allowed or proper green-arrow lengths. Section 6.1 demonstrates green-arrows of length 1, ½ and ¼ satisfy green-arrow laws of motion on a Cartesian grid of alternate one-way roads. Section 6.2 demonstrates green-arrows of length ½ and ¼ satisfy green-arrow laws of motion on a Cartesian grid of two-way roads and Section 6.3 summarizes what has been discovered up to this point. Section 2.8 implies results found for a Cartesian grid are applicable to existing plain road networks. Subsequently, the method illustrated in this section is computerized to find all permitted green-arrow lengths.

*6.1 Alternate one-way RGW-roads on Cartesian Road network.* Figure 17 demonstrates equal length green-arrows of length 1, ½ and ¼ satisfy green-arrow laws of motion on alternate one-way system of RGW-roads.

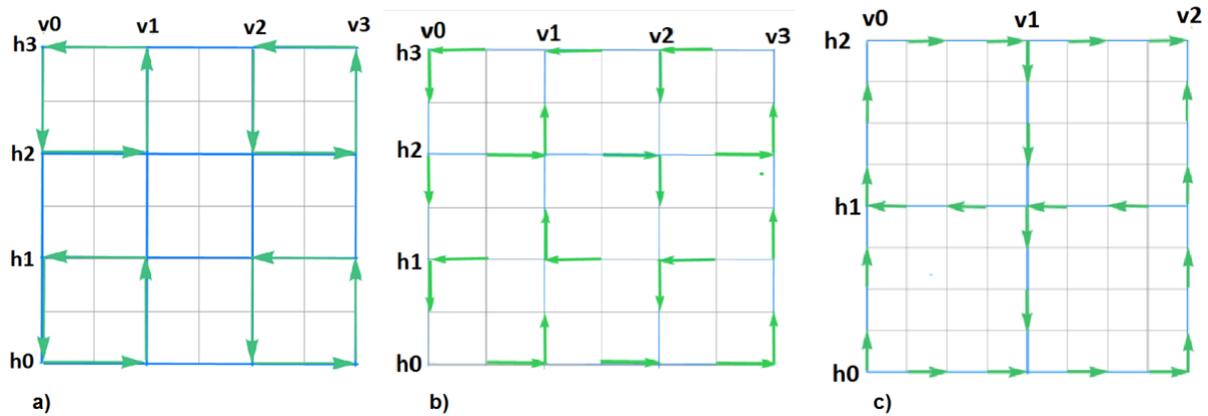

FIGURE 17. Demonstrating equal length green-arrows of length 1, ½ and ¼ satisfy green-arrow laws of motion on an alternate one-way Cartesian Road grid. a) Green-arrows of length 1. b) Green-arrows of length ½. c) Green-arrows of length ¼.

The reader can reproduce Figure 17 diagrams by placing a random green-arrow of specified length in one of the shown positions. The requirement that each intersection be fully utilized then determines the position of all other green-arrows at that instant of time. The reader can show each of the patterns shown in Figure 17 satisfy green-arrow laws of motion by advancing each arrow in time until the pattern reproduces itself and observing that green-arrows never intersect one another and each intersection is fully utilized at all time. Observe in Figure 17, that each green-arrow pattern is spatially and temporally periodic which implies these results applies to plain alternate one-way road networks of arbitrary size.

*6.2 Two-way RGW-roads on Cartesian Road network.* Figure 18 demonstrates equal length green-arrows of length ½ and ¼ satisfy green-arrow laws of motion on a two-way system of RGW-roads.

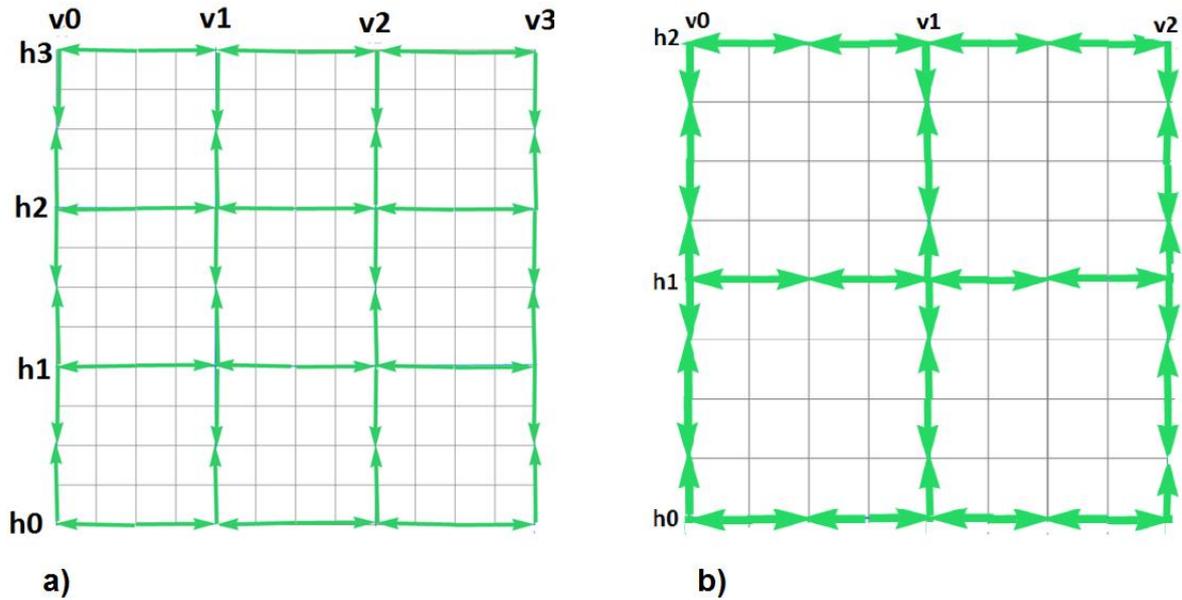

FIGURE 18. Demonstrating equal length green-arrows of length ½ and ¼ satisfy green-arrow laws of motion on a two-way Cartesian Road grid. a) Green-arrows of length ½. b) Green-arrows of length ¼.

One can reproduce Figure 18 by placing one green-arrow in a position shown in Figure 18. The requirement that each intersection be fully utilized results in a reproduction of Figure 18. One can show that Figure 18 satisfies green-arrow laws of motion by advancing each green-arrow until the pattern is reproduced and noting that green-arrows never intersect one another and fully utilize each intersection. Observe that each pattern in Figure 18 is spatially and temporally periodic which implies these results applies to plain two-way road networks of arbitrary size.

*6.3 Summary of what is now known*. For equal length green-arrows on a Cartesian alternate one-way road network: 1) the maximum green-arrow length is 2, 2) green-arrow lengths of 2, 1, ½ and ¼ satisfy green-arrow laws of motion, 3) some green-arrow lengths cannot satisfy green-arrow laws of motion on this road grid. For equal length green-arrows on a Cartesian two-way road network: 1) the maximum green-arrow length is 1, 2) green-arrow lengths of 1, ½ and ¼ satisfy green-arrow laws of motion, 3) some green-arrow lengths cannot satisfy green-arrow laws of motion on this road grid. For every case where equal length green-arrow laws of motion are satisfied on a Cartesian Road grid: 1) green-arrows are spatially and temporally periodic and 2) green-arrow lengths are rational numbers.

## 7. Unequal length green-arrows on Cartesian Road networks

Section 7.1 demonstrates that unequal green-arrows (Section 2.24) can satisfy green-arrow laws of motion on a Cartesian Road grid and notes that when this is possible only when the sum of the green-arrow lengths in orthogonal directions have discrete values. The discrete values are related to the discrete values of equal length (Section 2.23) green-arrows. It is also shown that with unequal green-arrows the singular point becomes a singular interval on roads with the longer length green-arrows and disappears on roads with the shorter length green-arrows. Sections 7.2, 7.3, 7.4 and 7.5 demonstrate that green-arrow laws of motion are satisfied on alternate one-way Cartesian Road grids when the length of unequal green-arrows sum to 4, 2, 1, ½ respectively. Although the results are demonstrated for alternate one-way Cartesian Road grids they apply to existing plain one-way road networks. Anisotropy degree is defined in Section 7.6.

*7.1 Unequal length green-arrows satisfy green-arrow laws of motion*. Figure 19 shows unequal green-arrow position at an instant of time on a Cartesian Road network.

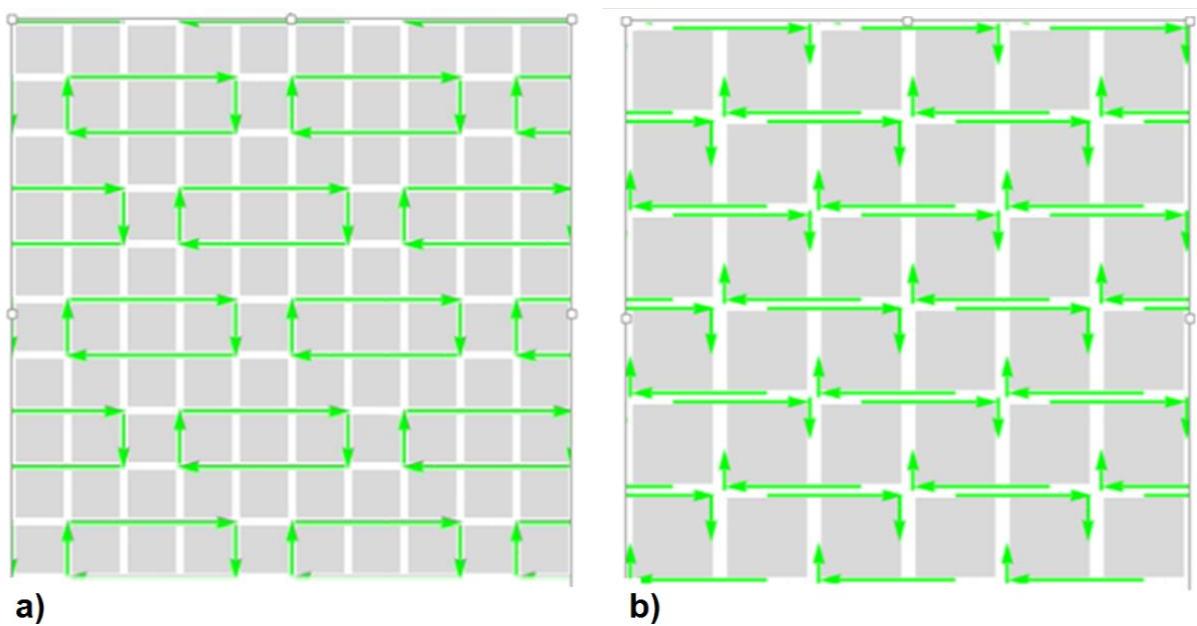

FIGURE 19. Unequal length green-arrow positions at an instant of time. a) Alternate one-way roads with east-west and north-south green-arrow lengths of 3 and 1. b) Two-way roads with east-west and north-south green-arrow lengths of 3/2 and ½.

In Figure 19 a) observe that in the north/south and east/west directions green-arrows are periodic in space with a period of 4 and for Figure 19 b) in both directions green-arrows are periodic in space with a period of 2. By graphically advancing green-arrows in Figure 19 a) by four Blocks it is determined that green-arrow positions fully utilize each intersection and that green-arrows never intersect one another. Similarly, by graphically advancing green-arrows in Figure 19 b) by two Blocks it is determined that green-arrows fully utilize each intersection and never intersect one another. Animations of Figure 19 ([video-4](video-4), [video-5](video-5)) demonstrate that green-arrows in Figure 19 satisfy green-arrow laws of motion.

Figure 19 is obtained from figures 9 and 11 by simultaneously lengthening east-west and shortening north-south green-arrows. Singular points exist at each mid-Block position in

Figure 11. One can ascertain, either from video-5 or by examining the movement of green-arrows implied by Figure 19 b), that singular points along north-south RGW-roads in Figure 19 b) have disappeared while the singular points along east-west roads have turned into singular intervals. These singular intervals are one of the disadvantages of meeting demand for anisotropic flow by using unequal green-arrows.

Two interesting observations. 1) For the alternate one-way road system of Figure 9, the sum of green-arrow lengths in the north-south and east-west direction is $2 + 2 = 4$; for the alternate one-way road system in Figure 19 a) the length of green-arrows in the north-south and east-west direction is $3 + 1 = 4$. 2) For the two-way road system of Figure 11, the sum of green-arrow length in the north-south and east-west direction is $1 + 1 = 2$; for the two-way road system in Figure 19 b) the length of green-arrows in the north-south and east-west direction is $3/2 + 1/2 = 2$.

*7.2 Proper alternate one-way green-arrow lengths sum to 4*. Section 7.1 showed that for an alternate one-way Cartesian Road network a green-arrow length of 3 in the east/west direction and a green-arrow length of 1 satisfy green-arrow laws of motion. In this section it is shown that for unequal length green-arrows on an alternate one-way Cartesian Road network the length of the individual green-arrows can have any positive length value (even irrational values) such that their sum is four. Figure 20 facilitates showing the validity of this result. In Figure 20, $\alpha$ and $\beta$ are green-arrow lengths in the north-south and east-west directions respectively. Equal length green-arrows means $\alpha = \beta$. The symbol $x$ is the length of the green-arrow on road $h4$ between $v1$ and $v2$, and each Block has length 1. Hence, the length of the green-arrow on $v1$ between $h3$ and $h4$ is 1 which implies the length of the green-arrow on $v1$ between $h4$ and $h5$ is $\alpha - 1$. Since all green-arrows travel at the same speed, the requirement that intersection $(h4, v1)$ be fully utilized implies the distance from intersection $(h4, v1)$ to the tip of the green-arrow on $h4$ is $\alpha - 1$. Since the distance between intersection $(h4, v1)$ and intersection $(h4, v2)$ is 1 it must be that

$$\alpha - 1 + x = 1$$

Figure 20 implies

$$\beta - x = 2$$

The last two equations imply

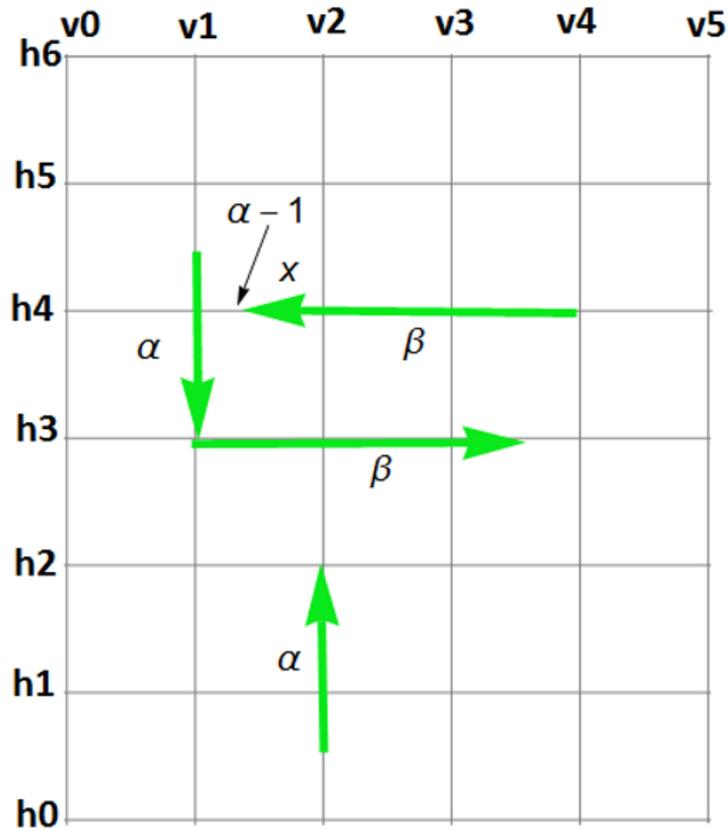

FIGURE 20. Demonstration that $\alpha + \beta = 4$ satisfies green-arrow laws of motion where $\alpha$ and $\beta$ are respectively green-arrow lengths in the north-south and east-west directions. Alternate one-way roads on the Cartesian roan network are labeled $h0, \ldots, v5$.

$$\alpha + \beta = 4 \tag{7}$$

When green-arrow lengths satisfy equation 7 and the green-arrows on $v1$ and $h4$ are placed as shown in Figure 20 and other green-arrows are placed so as to fully utilize each intersection then the green-arrows so placed will satisfy green-arrow laws of motion.

In Figure 9, $\alpha = \beta = 2$ illustrates the validity of (7). Figure 19 a) was obtained from Figure 9 by lengthening $\beta$ by 1 and shortening $\alpha$ by 1.

Up to this point allowed green-arrow lengths on alternate one-way roads are 2, 1, ½ and ¼, all rational lengths. We pose the question: does (7) imply lengths $\alpha$ and $\beta$ are rational? The answer is *no* as shown by the following example. Suppose green-arrow length in the north-south direction is $\alpha = 2 - \sqrt{2}$ and green-arrow length in the east-west direction $\beta = 2 + \sqrt{2}$. Then $\alpha$ and $\beta$ satisfy (7) but green-arrow lengths $\alpha$ and $\beta$ are irrational.

*7.3 Proper alternate one-way green-arrow lengths sum to 2.* Figure 21 demonstrates that green-arrow laws of motion are satisfied with unequal length green-arrows on an alternate one-way Cartesian Road network when the sum of the lengths is 2. In Figure 21, a northbound green-arrow of length $\alpha$ is entering intersection h1-v3 and an eastern heading wave of length $\beta$ has just passed intersection $(h1, v2)$. The requirement for maximum flow through intersection $(h1, v3)$ implies the distance from the head of the arrow traveling on $h1$ to intersection $(h1, v3)$ is $\alpha$. This implies $1 - \alpha$ is the length of the arrow on $h1$ between intersections $(h1, v2)$ and $(h1, v3)$. Since the distance between intersection h1-v1 and h1-v2 is one we conclude

$$\beta - (1 - \alpha) = 1 \ .$$

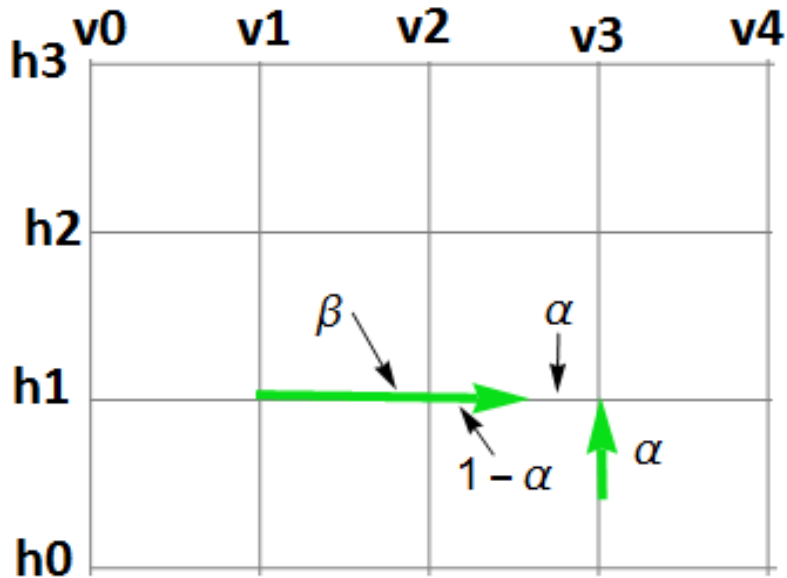

FIGURE 21. Demonstrating that $\alpha + \beta = 2$ satisfy green-arrow laws of motion where $\alpha$ and $\beta$ are respectively green-arrow lengths in the north-south and east-west directions. Alternate one-way roads on the Cartesian Road network are labeled $h0, ..., v4$.

This implies

$$\alpha + \beta = 2 \tag{8}$$

Here $\alpha$ and $\beta$ are any positive numbers, rational or irrational, that satisfy (8). Figure 17 a) is an example (here $\alpha = \beta = 1$) which demonstrates the validity of (8).

*7.4 Proper alternate one-way green-arrow lengths sum to 1*. Figure 22 demonstrates that green-arrow laws of motion are satisfied with unequal length green-arrows on an alternate one-way Cartesian Road network when the sum of the lengths is 1. Arterial roads $h0, ..., v3$ define Blocks in Figure 22. To facilitate the demonstration each block is filled with four squares. Green-arrows of length $\alpha$ and $\beta$ are shown in the figure. Clearly the length of wave $\alpha$ is less than ½ and the length of wave $\beta$ is greater than ½. To fully utilize intersection $(h1, v2)$, the distance from intersection$(h1, v2)$ to the tip of the arrow on $h1$ must be $\alpha$ as shown in

Figure 22. Let $x$ denote the portion of the green-arrow $\beta$ between point **P2** and intersection $(h1, v2)$. Then $x = 1/2 - \alpha$. This implies the length of the green-arrow on $h1$ between **P1** and **P2** is $\beta - x$. Figure 22 implies the distance between **P1** and **P2** is ½. Thus

$$\beta - (1/2 - \alpha) = 1/2$$

which simplifies to

$$\alpha + \beta = 1 \qquad (9)$$

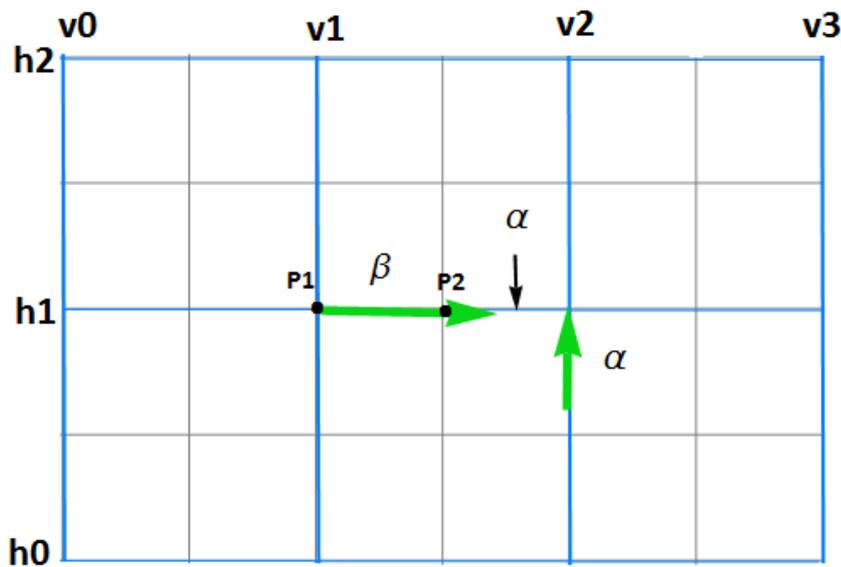

FIGURE 22. Demonstrating that $\alpha + \beta = 1$ satisfy green-arrow laws of motion. Alternate one-way roads on the Cartesian Road network are labeled $h0, \ldots, v3$.

Figure 17 b) which has $a = \beta = 1/2$ is an example which illustrates the validity of (9).

*7.5 Proper alternate one-way green-arrow lengths sum to ½.* Figure 23 demonstrates green-arrow laws of motion are satisfied with unequal length green-arrows on an alternate one-way Cartesian Road network providing their lengths sum to ½.

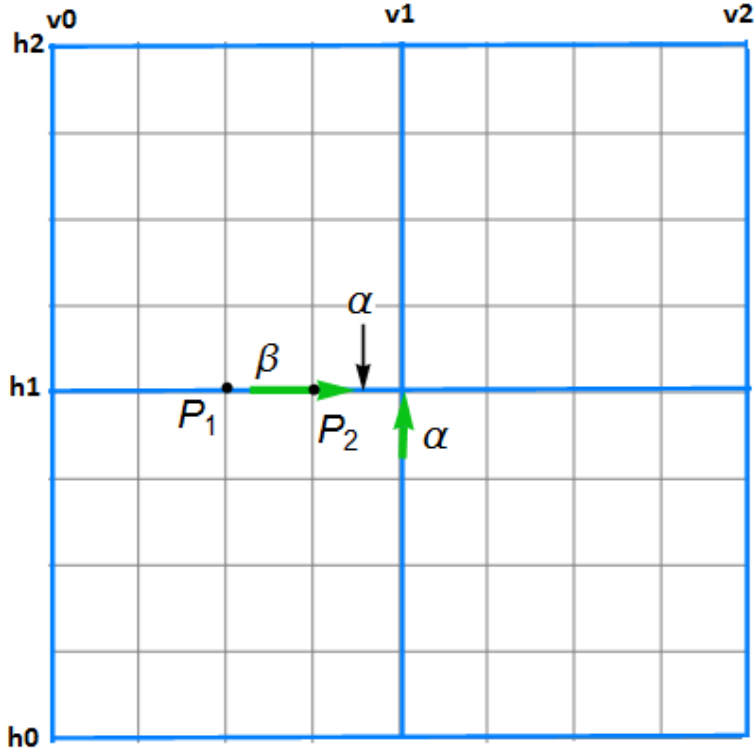

FIGURE 23. Demonstrating that $\alpha + b = 1/2$ satisfy green-arrow laws of motion. Alternate one-way streets are labelled $h0, \ldots, v2$.

In Figure 23, Blocks are bounded by arterial roads $h0, \ldots, v2$ and each Block is divided into 16 squares. Using an argument similar to that of Figure 22 yields the result

$$\alpha + \beta = 1/2 \tag{10}$$

Figure 17 c) which has $\alpha = \beta = 1/4$ can be converted to Figure 23 by suitably adjusting green-arrow lengths $\alpha$ and $\beta$ subjection to constraint (10).

*7.6 Anisotropy degree.* Define $\bar{\alpha}$ and $\bar{\beta}$ by

$$\bar{\alpha} \stackrel{\text{def}}{=} Max[\alpha, \beta], \quad \bar{\beta} \stackrel{\text{def}}{=} Min[\alpha, \beta]$$

Anisotropy degree $\gamma$ is then defined by

$$\gamma \stackrel{\text{def}}{=} \frac{\bar{\alpha}}{\bar{\beta}}$$

For isotropic flow ($\alpha = \beta$), and $\gamma = 1$. As $\gamma$ increases the singular interval increases which implies to get anisotropic flow using unequal length green-arrows it is desirable to keep $\gamma$ as small as possible.

## 8. A compilation of equal length proper green-arrow lengths.

Preceding sections showed that on a two-way or alternate one-way existing plain road network equal length green-arrows can be placed so as to satisfy green-arrow laws of motion but this can only be done for particular green-arrow lengths. The previous section showed that unequal length green-arrow patterns can be derived from equal length green-arrow patterns. The relationship of a Cartesian Road network to existing road networks is discussed briefly in Section 2.8 and more fully in [1]. The purpose of this section is to describe and use computer programs to generate all proper green-arrow lengths on two-way and alternate one-way Cartesian Road networks.

In Section 3 it was shown how the placement of one green-arrow at one instant of time determines the location of all green-arrows at all locations for all time. This section shows how to automate Section 3 results.

Sections 8.1 and 8.2 provide a compilation of all useful equal length proper green-arrow lengths on alternate one-way and two-way arterial roads respectively. Subsections provide a top-down description of the Mathematica code used to automatically draw green-arrow patterns that satisfy green-arrow laws of motion.

***8.1 Systematic search for proper green-arrow lengths on alternate one-way roads***. In developing code for automatically drawing green-arrow patterns that satisfy green-arrow laws of motion it is convenient to adopt the convention that green-arrows on the x-axis are westward bound and are shown at a moment of time when one green-arrow head is about to enter the origin (Figure 24). We also adopt the convention that green-arrows on the y-axis are southward bound and fully utilize the intersection at the origin.

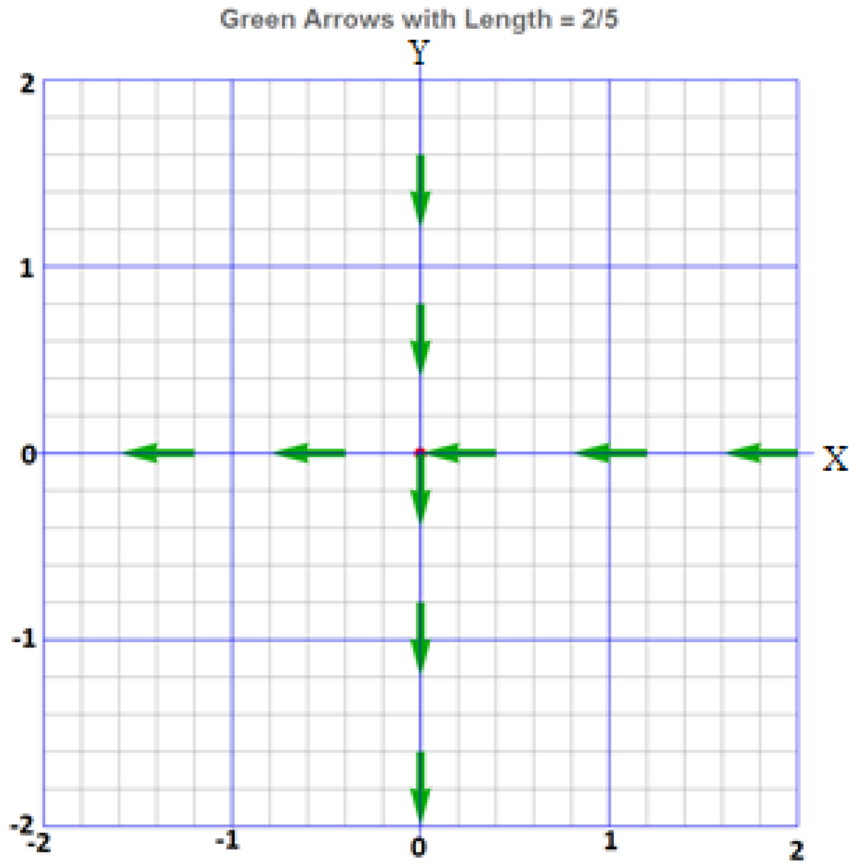

FIGURE 24. Development of program to find proper green-arrow lengths.

In Figure 24 blue lines indicate alternate one-way roads, green-arrows are 2/5 long and grey lines are drawn for convenience to facilitate discussion and to enable the reader to ascertain green-arrows are indeed 2/5 units long. The condition that Figure 24 roads are alternate one-way and the condition that the y-axis road is southbound implies roads defined by $x = -1, 1$ are northbound while roads defined by $x = -2, 0, 2$ are southbound. Similarly, roads defined by $y = -2, 0, 2$ are westbound and roads defined by $y = -1, 1$ are eastbound.

*8.1.1. Computer drawn green-arrow patterns, and compilation of proper equal length green-arrow lengths.* Using the same procedure used to construct Figure 10, every green-arrow on the x-axis in Figure 24 determines one north or southbound green-arrow for each integer value of $x$ and every green-arrow along the y-axis determines the location of one east or westbound green-arrow. For example, a northbound green-arrow with length equal to 2/5 and arrow head tip at $(1, -1/5)$ is needed to fully utilize intersection $(0, 1)$. All remaining green-arrows in Figure 24 are determined by the isotropic flow condition which implies the space between green-arrows equals green-arrow length. Computer inputs, at the top of Figure 25, initiate calculations which complete Figure 24.

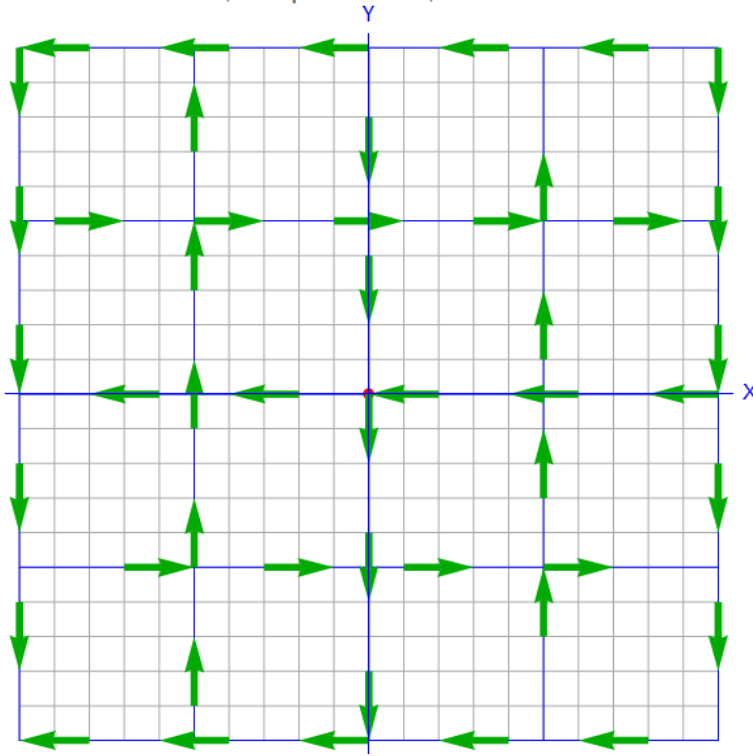

FIGURE 25. Computer generated equal length green-arrow pattern with length 2/5.

Code which generated Figure 25 is exhibited in Figure 26. Comments in the code are designated with a matching pair of parenthesis and asterisk (* Comment *). Lines are numbered $Lx$ where the line number is $x$. Line 1 creates the graph shown in Figure 25 without green-arrows. Lines $2-5$ generate instructions for drawing all north, south, east and westbound green-arrows. Line 6 creates the graph according to instructions stored in **temp1**, ..., **temp4**. Line 7 displays the graphics stored in **g0** and **gArrows**. Code in Figure 23 made use of several subroutines **SArrowP**, **MatchColS**, etc. A complete listing of code used to generate Figure 25, and a more detailed explanation for how the code works is given in Appendix B.

In Figure 15 it was shown that 3/2 is not a proper green-arrow length. Figure 27 illustrates how Figure 26 code deals with this case.

```
AltOneWayIsoGen[] := Module[{temp1, temp2, temp3, temp4, g0, gArrows},
   (* Program automatically draws all green arrows *)
   (* Program intended to confirm allowed green arrow lengths *)
   g0 = MGraph[];  (* L1 Graph without arrows *)
   temp1 = Flatten[ NArrowP /@ MatchColN[]]; (* L2 *)
   (* NArrowP operates on output of MatchColN which produces instructions
    for drawing all northbound arrows *)
   (* Flatten gets rid of not needed braces *)
   temp2 = Flatten[ SArrowP /@ MatchColS[]]; (* L3 *)
   (* See comments for temp1  *)
   temp3 = Flatten[ EArrowP /@ MatchColE[]]; (* L4 *)
   (* See comments for temp1 *)
   temp4 = Flatten[ WArrowP /@ MatchColW[]]; (* L5 *)
   (* See comments for temp1 *)
   gArrows = Graphics[{dgreen, Thickness[Lthick], Arrowheads[Asize],
      temp1, temp2, temp3, temp4}]; (* L6 *)
   (* graph of all arrows *)
   Show[g0, gArrows] (* L7 *)]
```

FIGURE 26. Mathematica code used to generate Figures 25 and 27.

FIGURE 27. Code exhibited in Figure 26 demonstrates that 3/2 is not an allowed green-arrow length on alternate one-way roads.

Using Figure 26 code we systematically find all proper green-arrow lengths on alternate one-way roads. The task is facilitated by the result, shown to be true in Figure 13, that 2 is the longest possible green-arrow length on an alternate one-way road system. The task is also facilitated by the following conjectures, which are suggested by the geometry of the repeating green-arrow patterns in an alternate one-way RGW-road network.

**Conjecture 1**. On a Cartesian alternate one-way RGW-road network with equal length green-arrows, the green-arrow length must be a rational number.

**Conjecture 2**. On a Cartesian alternate one-way RGW-road network with equal length green-arrows, the green-arrow length $\mathbb{L}_{EAOW}$ cannot satisfy $1 < \mathbb{L}_{EAOW} < 2$.

Our systematic procedure for finding proper green-arrow lengths is to try integer numerators such that green-arrow length never exceeded 2 and to choose integer denominators that did not exceed 8. Table 1 gives all allowed equal length green-arrow lengths on alternate one-way roads between 2 and ¼. We felt green-arrow lengths smaller than ¼ are of little interest because shorter green-arrows imply more yellow and more all-red time spent clearing intersections. Our systematic search for proper green-arrow lengths and the program for finding them can be downloaded from Appendix B. The results of our systematic search are shown in Table 1.

Table 1. Proper green-arrow lengths for equal length green-arrows on alternate one-way roads.

| 2 | 1 | 2/3 | 1/2 | 2/5 | 1/3 | 2/7 | 1/4 |
|---|---|-----|-----|-----|-----|-----|-----|

Proper equal length green-arrow lengths in Table 1 are sorted with the largest proper lengths first since it is anticipated these lengths will be most important in applications. Based on our numerical investigation we hypothesize that *all* equal length proper green-arrow lengths on alternate one-way roads are given by the expression

$$\mathbb{L}_{EAOW} = \frac{2}{n}, \quad n = 1, 2, 3, \dots, \infty \tag{11}$$

The *EAOW* subscript is short for Equal length Alternate One-Way. In Appendix A, (11) is analytically shown to be valid, which we state as Proposition 1.

**Proposition 1**. On a Cartesian alternate one-way RGW-road network with equal length green-arrows, the green-arrow length must satisfy (11). Conversely, if the green-arrow length satisfies (11), there exists a corresponding Cartesian alternate one-way RGW-road network.

Mathematica code shown in Figure 28 demonstrates (11) and describes computer calculation results, exhibited in Table 1.

```
temp = Table[2 / n, {n, 1, 8}]  (* L1 *)
```

$$\left\{2, 1, \frac{2}{3}, \frac{1}{2}, \frac{2}{5}, \frac{1}{3}, \frac{2}{7}, \frac{1}{4}\right\}$$

FIGURE 28. Mathematica code demonstrates (11) and Table 1 results are consistent.

The first line in Figure 28 computes values of $2/n$ for $n = 1, \dots, 8$ which reproduces values found in Table 1.

A second proposition is formulated in terms of the spatial period $\xi$. Figure 25 and Table 1 show that $2/5$ is a proper green-arrow length. The spatial period in Figure 25 is four because four is the smallest distance needed to advance in the direction of the arrow before the green-arrow pattern repeats exactly. This is readily seen by examining the points $(2,0)$ and $(-2,0)$ in Figure 25. At the point $\{2,0\}$ there is a green-arrow pointing south entering the intersection and another green-arrow pointing west leaving the intersection and the same is true for the point $\{-2,0\}$. This implies the proper length 2/5 has a spatial period of 4 as indicated in. Table 2 was produced by examining green-arrow patterns exhibited in Appendix B and noting the relationship between proper equal length green-arrow lengths $\mathbb{L}_{EAOW}$ and spatial period $\xi$ on alternate one-way roads. Table 2 suggests Proposition 2.

Table 2. Spatial periods and equal length proper green-arrow lengths on alternate-one-way roads.

| $\mathbb{L}_{EAOW}$ | 2 | 1 | 2/3 | 1/2 | 2/5 | 1/3 | 2/7 | 1/4 |
|---|---|---|---|---|---|---|---|---|
| $n$ | 1 | 2 | 3 | 4 | 5 | 6 | 7 | 8 |
| $\xi$ | 4 | 2 | 4 | 2 | 4 | 2 | 4 | 2 |

**Proposition 2**. The spatial period of the pattern of green-arrows in a Cartesian alternate one-way RGW-road network with equal length green-arrows is either 2 or 4. In particular, when the green-arrow length is given by (11), the spatial period is 2 if $n$ is even and 4 if $n$ is odd.

Proposition 2, shown to be true in Appendix A, has an interesting geometric interpretation. Realize that green-arrow graphs shown in Figure 25 and in Appendix B show only a small portion of the Cartesian grid which covers the entire $x - y$ plane. The equal length green-arrow requirement and green-arrow laws of motion imply that for green-arrows with $\xi = 2$ any $2 \times 2$ portion from a green-arrow diagram can be used to tile the entire Cartesian grid and for green-arrows with $\xi = 4$ any $4 \times 4$ portion from a green-arrow diagram can be used to tile the entire Cartesian grid. Any equal length proper green-arrow diagram with a spatial period of 2 also has a spatial period of 4. Thus, all equal length proper green-arrow diagrams have a spatial period of 4.

*8.2 Systematic search for equal length proper green-arrow lengths on two-way roads*. This section describes proper equal length green-arrow lengths on a Cartesian network of two-way roads. The Mathematica code used to discover these values is described in Section 8.2.1. Also given is a compilation of equal length proper green-arrow lengths on two-way roads.

*8.2.1 Code for drawing green-arrows on two-way roads and compilation of proper green-arrow lengths*. In Section 8.1.1 the instant of time is chosen so that on the x-axis green-arrows are entering the origin and the same convention is used for automating green-arrow placement on two-way roads. The condition that the intersection at the origin be fully utilized implies that when eastbound green-arrow enter this intersection north and southbound arrows leave the intersection as illustrated in Figure 29. Green-arrows with length ½ on the x and y axes fully utilize each RGW-intersection on the coordinate axes.

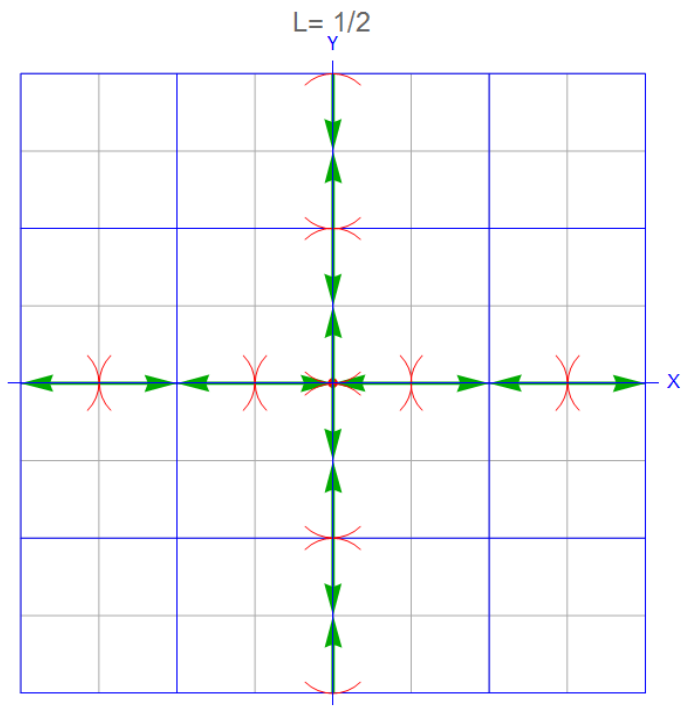

FIGURE 29. Green-arrow with length ½ on the x- and y-axes. Red arcs mark the end of green-arrows.

With two-way traffic, green-arrows can overlap as indicated in Figure 11 or their tails can touch one another as in Figure 29. The tail of each green-arrow is marked by an arc segment of a red circle centred on its arrow head to facilitate image understanding.

The following discussion demonstrates the green-arrows shown in Figure 29 and the requirement that green-arrows obey green-arrow laws of motion determine the placement of all the north, south, east and westbound green-arrows. Green-arrows at intersections along the x-axis are entering each intersection and the condition for maximum flow indicates north and southbound arrows must have just left these intersections. Similarly green-arrows at intersections on the y-axis have just left the intersection so east and westbound green-arrows must be entering these intersections. Using green-arrow spatial periodicity, Figure 29 is completed to get Figure 30.

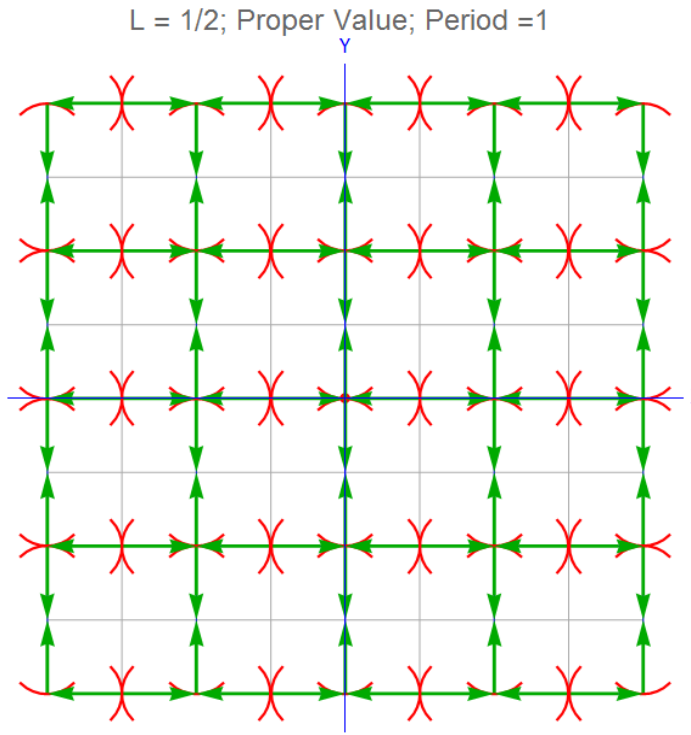

FIGURE. 30. Computer generated equal length green-arrow pattern with length equal to ½ on a Cartesian grid of arterial roads. Red circle arc segments are centred on green-arrow heads and denote the end of green-arrows.

Code which generated Figure 30 is exhibited in Figure 31. In Figure 31, Line 1 creates the graph green-arrows are drawn on. Lines $2-5$ generate instructions for drawing all the north, south, east and westbound arrows. Line 6 creates the graph according to instructions stored in **temp1**, ... , **temp4** . Line 7 displays the graphics stored in **g0** and **gArrows.** Code in Figure 31 made use of several subroutines **NArrowP**, **MatchColNT**, etc. A complete listing of code used to generate Figure 31, and a more detailed explanation for how the code works is given in Appendix B.

```
TwoWayIsoGen[] :=
 Module[{temp1, temp2, temp3, temp4, g0, gn, gs, gArrows},
  (* Program automatically draws all green arrows *)
  (* Program intended to confirm allowed green arrow lengths *)
  g0 = MGraph[];  (* L1 Graph without arrows *)
  temp1 = NArrowP /@ MatchColNT[];  (* L2 *)
  (* NArrowP operates on output of MatchColNT which produces
   instructions for drawing all northbound arrows *)
  temp2 = SArrowP /@ MatchColST[];  (* L3 *)
  (* See comments for temp1  *)
  temp3 = EArrowP /@ MatchColET[];  (* L4 *)
  (* See comments for temp1 *)
  temp4 = WArrowP /@ MatchColWT[];  (* L5 *)
  (*  See comments for temp1 *)
  gArrows =
   Graphics[{dgreen, Thickness[Lthick], Arrowheads[Asize],
     temp1, temp2, temp3, temp4}];  (* L6 *)
  Show[g0, gArrows]  (* L7 graph grid and all arrows *)]
```

FIGURE 31. Mathematica code used to generate Figure 30.

Figure 16, generated by the code in Figure 32, demonstrates that 3/2 is not a proper green-arrow length on a Cartesian two-way road network.

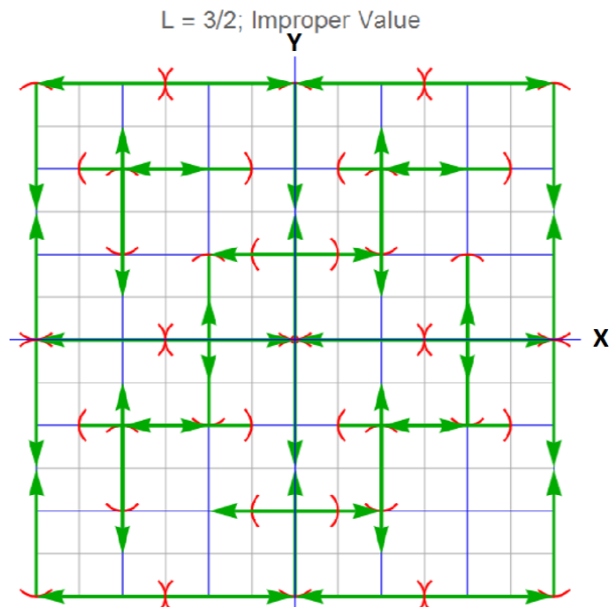

FIGURE 32. On two-way roads equal length green-arrow length 3/2 is an improper length. At the intersection $\{-2,-1\}$ a green-arrow heading north between $\{-2,-1/2\}$ and $\{-2,-2\}$ intersects a green-arrow heading east between $\{-1,-1\}$ and $\{-1,-5/2\}$.

Similar to conjectures 1 and 2 for Cartesian alternate one-way RGW-road networks, we make the following conjectures for two-way RGW-road networks.

**Conjecture 3**. On a Cartesian two-way RGW-road network with equal length green-arrows, the green-arrow length must be a rational number.

**Conjecture 4.** On a Cartesian two-way RGW-road network with equal length green-arrows, the green-arrow length cannot exceed 1.

Utilizing conjectures 3 and 4, and the result (Figure 14) that on a two-way Cartesian Road grid equal length green-arrow lengths greater than one are not possible, we systematically explored all rational green-arrow length between 1 and 1/8 inclusive. Proper and improper green-arrow lengths were found using the procedure illustrated in figures 30 and 32. Results are shown in Table 3 with proper green-arrow length sorted according to length. Here $L_{ETW}$ represents **E**qual length proper green-arrow lengths on a Cartesian **T**wo-**W**ay Road network.

*Table 3. Proper equal length green-arrow lengths on a Cartesian two-way road network.*

| $L_{ETW}$ | 1 | 1/2 | 1/3 | 1/4 | 1/5 | 1/6 | 1/7 | 1/8 |
|---|---|---|---|---|---|---|---|---|

Detailed calculations and the Mathematica code used to generate Table 3 are given in Appendix B.

Based on the numerical investigation summarized in Table 3 we hypothesize: on a Cartesian two-way road network proper equal length green-arrow lengths are given by the expression

$$\mathbb{L}_{ETW} = \frac{1}{n}, \quad n = 1,2,3,\ldots,\infty \tag{12}$$

In Appendix A, (12) is analytically shown to be valid, which we state as Proposition 3.

**Proposition 3**. On a Cartesian two-way RGW-road network with equal length green-arrows, the green-arrow length must satisfy (12). Conversely, if the green-arrow length satisfies (12), there exists a corresponding Cartesian two-way RGW-road network.

In Figure 30 the green-arrow pattern is unchanged when each green-arrow is advanced by one Block. We denote this by saying the green-arrow pattern of Figure 30 has a spatial period $\xi = 1$. Using results shown in Appendix B, the spatial period for equal length proper green-arrow lengths is shown in Table 4.

Table 4. Spatial period on isotropic flow Cartesian two-way road networks.

| $L_{ETW}$ | 1 | 1/2 | 1/3 | 1/4 | 1/5 | 1/6 | 1/7 | 1/8 |
|---|---|---|---|---|---|---|---|---|
| $n$ | 1 | 2 | 3 | 4 | 5 | 6 | 7 | 8 |
| $\xi$ | 2 | 1 | 2 | 1 | 2 | 1 | 2 | 1 |

Examination of Table 4 suggests Proposition 4.

**Proposition 4**. The spatial period $\xi$ of the pattern of green-arrows in a Cartesian two-way RGW-road network with equal length green-arrows is either 1 or 2 and

$$\xi(n) = \begin{cases} 1 & \text{When } n \text{ is even} \\ 2 & \text{When } n \text{ is odd} \end{cases}, \quad n = 1, 2, \ldots \infty \quad (13)$$

Equations (12), and (13), found by computer calculations, are analytically derived in Appendix A.

## 9. Left-turn- and reduced-flow-arrows.

This section describes left-turn-arrows which are useful for creating progressions with anisotropic flow when maximum flow is not needed. Reduced-flow-arrows are always needed when left-turn-arrows are used. Section 9.1 describes how left-turn-arrows relate to work described in preceding sections. Left-turn-arrow notation, how it is described on the RTFD and how a motorist relates to RTFD information is described in Section 9.2. Some methods for using left-turn-arrows are described in Section 9.3.

*9.1 Reprise*. Section 6 described how different length green-arrows can satisfy green-arrow laws of motion to achieve maximum throughput uninterrupted flow on alternate one-way or two-way road networks that are topologically equivalent to a Cartesian Road network. In Section 4 it was found that the maximum green-arrow length on alternate one-way (two-way) road networks is two (one). When a two-way road network is operating at maximum flow the only way to make left turns is by using left turn arounds (Section 2.5). It has been shown [1] that on an existing suburban two-way road, equal length green-arrows simultaneously achieved maximum isotropic throughput uninterrupted flow in two opposing directions. However, the task of a traffic engineer is to time traffic signals on a road network so that it optimally meets traffic flow demands which is rarely isotropic.

Maximal throughput with unequal length green-arrows is described in Section 7 but need to be used cautiously because of the singular interval (sections 2.15 and 7.1). In timing traffic signals for anisotropic flow, the anisotropy factor $\gamma$ should be minimized subject to the constraint that it accommodates the anisotropic flow demand because doing so minimizes the singular interval.

Green-arrow laws of motion which ensure maximum potential flow in the forward direction imply that direct left turns cannot be made between RGW-roads. At times where maximum flow is needed in all directions, left-turn-arounds (Section 2.5) are used. Left-turn-arrows allow the more convenient direct left turns to be made under two conditions: 1) off peak flow or 2) when there is a substantially different flow demand in opposite directions.

*9.2 Left-turn-arrow notation and RTFD representation*. This section describes left-turn-arrows and reduced-flow-arrows by describing vehicle behaviour in different portions of these arrows. An important rule: left-turn-arrows and reduced-flow-arrows obey green-arrow laws of motion, i.e. they move like green-arrows. When vehicles in the green portion of a left-turn-arrow come to an RGW-intersection, the vehicles can continue in the forward direction or make a right turn. When vehicles in the blue portion of a Figure 33 d) left-turn arrow enter a RGW-intersection they can always make a left or right turn without stopping and for a Figure 33 c) left-turn-arrows they can also continue in the forward direction without stopping as well. The blue portion of a left-turn-arrow can have the blue portion at the beginning or end or at the beginning and end of a left-turn-arrow.

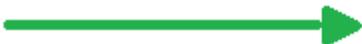

FIGURE 33. Illustrating green-, left-turn- and reduced-flow-arrows.

When the red vertical bars in c), e) and f) of Figure 33 enter an intersection they turn the traffic signal red on the side of the road they point to and when the blue portion of a left-turn-arrow enters an intersection it turns on a traffic signal indicating a direct left turn is possible for vehicles in the blue portion of a left-turn-arrow. The red bar pointing to both sides of the road in Figure 33 d) indicates that red lights are turned on for both sides of the road and that motorists can make a left or right turn without stopping for a traffic light but cannot continue in the forward direction until the traffic light turns green.

The red part of a reduced-flow-arrow turns on a red traffic signal on the side of the road the reduced-flow-arrow is located. The behaviour of traffic signals is dictated by the requirements that vehicles in the green portion of a green-arrow can travel forward through an intersection and vehicles in the blue portion of a left-turn-arrow can always make a right or left turn without stopping for a traffic signal but sometimes have to stop for a traffic signal if they want to go forward. As in Figure 1, red dots can be placed on the green part of the arrow shaft to indicate when and for how long pedestrians have a walk signal. Reduced-flow-arrows g), h) and i) are used respectively with left-turn-arrows c), e), and f).

Left-turn- and reduced-flow-arrows are intended to help traffic engineers visualize how to coordinate traffic signals to obtain the proper balance between demand for flow in the forward and left turn directions. Since left-turn- and reduced-flow-arrows move like green-arrows, when properly placed, they are guaranteed not to collide. Green-arrows fully utilize each intersection in the forward direction to gain maximum throughput. Left-turn and reduced-flow-arrows also fully utilize each intersection to achieve the convenient direct left turn but do so by sacrificing throughput.

Figure 34 illustrates how green-, left-turn-, and reduced-flow-arrows are represented in the RTFD. In the RTFD green-, left-turn, and reduced-flow-arrows all point to the top of the display which is the direction of motion. Here they are shown horizontally to create a compact figure. Arrow heads are not needed in the RTFD because all arrows always move toward the top of the display. In Figure 33 reduced-flow-arrows g), h), and i) point in the opposite direction of left-turn-arrows c), e), and f) since reduced-flow-arrows always move in the opposite direction to their corresponding left-turn-arrows. In Figure 34 reduced-flow-arrows g), h), and i) point in the same direction as left-turn-arrows to make it easier to make the connection between arrow representation and RTFD display of arrow representation. Yellow is shown in the Figure 34 a) RTFD representation even though it is omitted in its green-arrow representation because it is important for the motorist to know he/she will encounter a yellow light at the next signalized intersection unless they get out of the yellow zone. To simplify traffic signal timing design diagrams, those details are not included in the arrow representation. Although not shown in Figure 34, red dots on the green portion of a left-turn-arrow indicate to the motorist when and for how long pedestrians have a walk signal.

| Arrow Representation | Name | RTFD Display |
|---|---|---|
| a) → | Simplified green-arrow | |
| b) → | Detailed green-arrow | |
| c) → | Left-turn-arrow head forward | |
| d) → | Left-turn-arrow head stop | |
| e) → | Left-turn-arrow tail | |
| f) → | Left-turn-arrow head & tail | |
| g) → | Reduced-flow-arrow head | |
| h) → | Reduced-flow-arrow tail | |
| i) → | Reduced-flow-arrow head & tail | |

FIGURE 34. Arrow representation, name and RTFD display.

The use of the RTFD for a motorist is straightforward. As long as the motorist remains in the green portion of a green-arrow he/she will make every traffic signal on the road the motorist is currently on and will be able to make a right turn without stopping for a traffic signal. As long as the motorist remains in the blue portion of a left-turn-arrow the motorist will be able to make a left or right turn without stopping for a traffic signal and depending on whether there is a red bar at the head of the blue rectangle (Figure 34 c or d) the motorist will be able to go forward (make the light) or be forced to stop. The left-turn-arrow illustrated in Figure 34 f) is useful late at night when there is little traffic flow demand. The ability to make direct left turns without having to stop for a traffic signal provided by left-turn-arrows is bought at the expense of reduced traffic flow in the opposite direction. When there is little need for maximum flow in the opposite direction this is a reasonable tradeoff.

*9.3 Use of reduced-flow and left-turn-arrows*. Figure 35 illustrates the use of reduced-flow- and left-turn-arrows on a Cartesian grid of two-way roads. There is no need for left-turn-arrows or reduced-flow-arrows on alternate one-way roads. Examination of this figure shows that although reduced-flow- and left-turn-arrows move like equal length green-arrows they can be used to address anisotropic flow demand but do so at the expense of maximum total throughput.

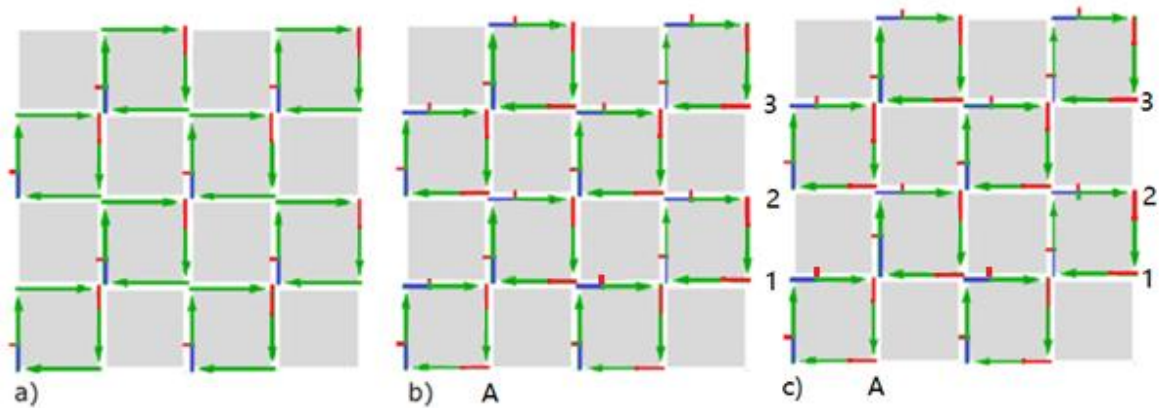

FIGURE 35. Illustrating use of reduced-flow- and left-turn-arrows. a) illustrates maximum throughput in east/west direction and maximum throughput in the northern direction with direct left turns in that direction at the expense of throughput in the southern direction. b) illustrates maximum throughput in the eastern and norhtern direction with direct left turns in these directions. c) is the same as b) with enhanced direct left turn capability on road A between roads 1 and 2.

The advantage of Figure 35 b) over Figure 35 a) was bought by sacrificing throughput in the southern and and western directions. Figure 35 c) is the same as Figure 35 b) with enhanced direct left turn capability on road A between roads 1 and 2.

Figure 35 shows how to meet anisotropic flow demand using reduced-flow- and left-turn-arrows but the figure does not show how to to use left-turn-arrows simultaneously in opposite directions. Figure 36 shows how to use left-turn-arrows in opposite directions. When left-turn-arrows are used in opposite directions vehicles in the blue portion of the left-turn-arrow can make a left or right turn by slowing down without stopping but are stopped by a red light if they want to go forward.

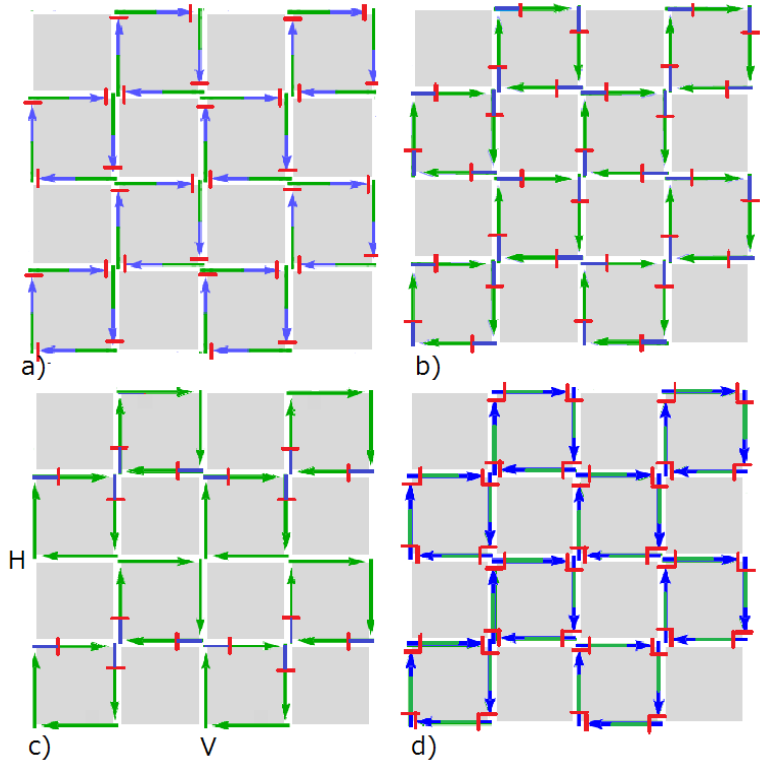

FIGURE 36. The use of left-turn-arrow without the use of reduced-flow-arrows. The blue portion of the left-turn-arrow is at the head of the arrow in a) and is at the tail of the left-turn-arrow in b) and and c). Figure 36 b) has left-turn-arrows on all roads but Figure 36 c) has full flow green-arrows on roads V and H. The use of left-turn-arrows with a blue portion at the head and tail is illustrated in Figure 36 d).

In Figure 36 motorists in the green portion of the left-turn-arrow make every traffic signal regardless of whether they are heading north, south, east or west. Here vehicles in the blue portion of the left-turn-arrow make left or right turns without stopping for a traffic signal.

For all diagrams in Figure 36 an ability to make direct left turns was achieved at the expense of forward flow throughput. Figures 36 a) and b) are intended for use when traffic flow demand is sufficiently low that direct left turns can be made from all roads. Because of full flow roads H and V in Figure 36 c) this figure is intended for use when there is too high a traffic flow demand to use Figures 36 a) and b). Figure 36 d) is intended for use when traffic flow demand is so low that direct left turns can be made from either the front or rear of a left-turn-arrow.

Figures 35 and 36 illustrate the use of left-turn-arrows based on isotropic flow green-arrows. Figure 37 shows the use of left-turn-arrows based on an anisotropic flow green-arrow diagram.

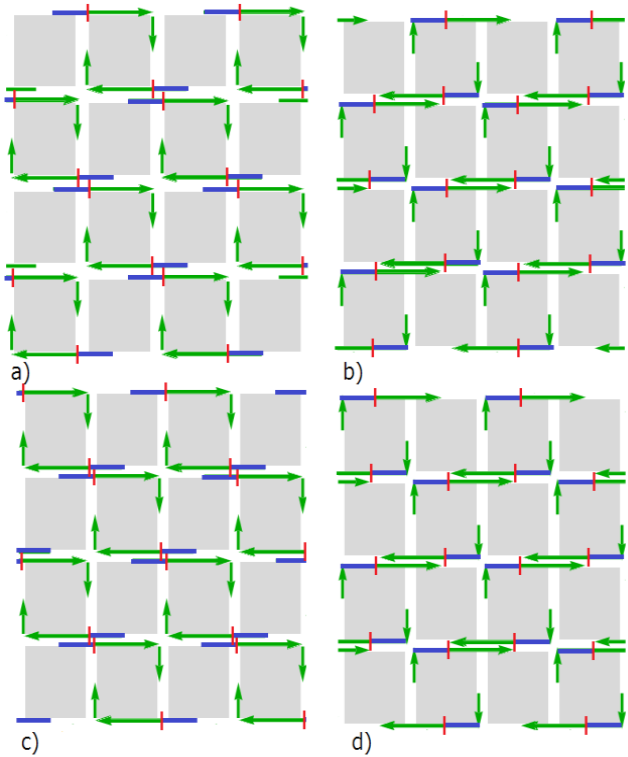

FIGURE 37. The use of left-turn-arrows on two-way roads based with left-turn-arrow length 3/2 in the east/west direction and green-arrows of length 1/2 in the north/south direction. a) Initial placement of left-turn-arrows. b) Left-turn-arrows have advanced ½ block from a). c) Left-turn-arrows have advanced 1 block from a). d) Left-turn-arrows have advanced 1 ½ blocks from a).

In Figure 37 green- and left-turn-arrows are shown at successive instants of time to demonstrate that each intersection is fully utilized. This was done to convince the reader that the configuration shown in a) will fully utilize each intersection although that is not really necessary since it was already shown in Figure 19 b). In Figure 19 b) each intersection is utilized fully to accommodate flow in the forward direction whereas Figure 37 utilizes each intersection fully but sacrifices flow in the forward direction to accommodate a direct left-turn capability.

Figure 38 shows how to use left-turn-arrows to get maximum flow toward or away from a city using left-turn-arrows.

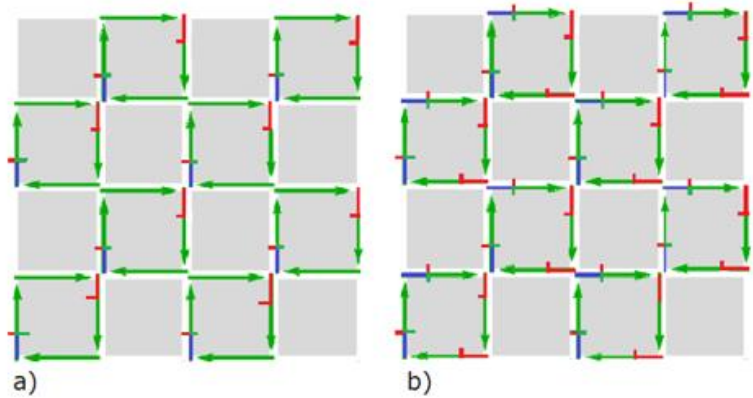

FIGURE 38. Maximum potential vehicle flow toward or away from a city using left-turn-arrows. a) Shows how to have maximum flow capacity in east, west and north directions with no-stop left turn capability in the north direction. b) Shows how to have maximum flow capacity in the east and north directions with no-stop left turns in the east and north directions.

## 10. Results, applications and conclusions

**Results and discussion**. The green-arrow laws of motion (Section 2.2) allow maximum throughput arbitrarily long traffic signal progressions to be found [1] using a new paradigm, i.e. move green-arrows to get uninterrupted flow while simultaneously utilizing each RGW-intersection in the forward direction to the fullest extent. This approach, modeled after Euclid's approach to plane geometry, uses postulates (green-arrow laws of motion) to achieve its end (maximum throughput and arbitrarily long progressions).

Left-turn and reduced-flow-arrows and their RTFD representation have been described. The notion of left-turn and reduced-flow-arrows, which move like green-arrows, is a new concept. A notation for left-turn and reduced-flow-arrows has been invented (Section 9). Left-turn and reduced-flow-arrows can be used to meet anisotropic flow demand using green-arrow laws of motion (Section 9). Left-turn and reduced-flow-arrows enable one to meet anisotropic flow demand while avoiding the singular interval inherent in the use of unequal length green-arrows.

Although cloverleaf left turns on highways/freeways have been used for many years, the adoption of this concept to get uninterrupted maximum flow in the forward direction on arterial roads is new (Section 2.5).

Green-arrow laws of motion and the mathematical consequence that on existing two-way arterial roads green-arrows move with variable speed, motivated the development of the RTFD (Section 2.4). This device enables: 1) maximum throughput arbitrarily long progressions on RGW-roads 2) enables uninterrupted flow for motorists on arterial roads and streets, and 3) enables a new class of roads termed RGW-roads that provide uninterrupted flow but which are much less expensive than highways/freeways [1]. RTFD capabilities, implemented as a smartphone application, is new.

The approach used to get maximum throughput uninterrupted flow implemented in this paper is fundamentally different from the usual approach to obtaining progressions. Typically, the traffic engineer defines a function and then uses a program to vary the input parameters with the goal of either minimizing or maximizing the defined function. The approach used here is deductive. Green-arrow laws of motion are the postulates. Euclidean-like geometric reasoning yield traffic signal timings with maximum bandwidth/throughput providing uninterrupted flow in four different directions in a grid network, which we call north, south, east and west. With the approach described here, the traffic engineer decides where RGW-nodes are to be placed [1]. Once the location of RGW-nodes and the cycle length are decided then the calculation of each traffic signal state calculated by the RGW-TLP program, but the calculations are simple enough to be done with a hand calculator. To get arbitrarily long progressions with maximum throughput analytical results, derived in this paper, are used instead of intricate computer calculations.

Several new mathematical results follow from the green-arrow laws of motion.

1) Although results 2 – 10, described below, all refer to alternate one-way and/or two-way Cartesian road grids, the discussion in Section 2.8 and [1] imply these results are applicable to any road network that is topologically equivalent to a Cartesian Road network.

2) Isotropic flow using green-arrows of equal length, traveling on a two-way Cartesian road grid result in a singular point at the mid-Block location (Figure 11 or video-3) and that for green-arrows of unequal length, the singular point becomes a singular interval (Figure 19 b or video-5) that includes the mid-Block position.

3) The placement of a single green-arrow at one instant of time on an alternate one-way or two-way Cartesian road grid determines the placement of every other green-arrow on that road grid at that instant of time and for all previous and future instants of time (Sections 3, and A.1.5). This result faacilitated the construction of programs for finding all proper equal length green-arrows (Section 8).

4) Section 8 computer calculations made use of conjectures 1 and 3 that on one- or two-way RGW-roads equal length green-arrows must have rational lengths. These proposions are shown to be valid in Sections A.1.6 and A.2.1.

5) Section 8 computer calculations made use of Proposition 2: the spatial period of equal length green-arrows on Cartesian alternate one-way roads always have spatial periods of either 2 or 4 which implied calculations done in Section 8 could be applied to a single intersection or any number of intersections. Proposition 2 is analytically shown to be true in Appendix A.

6) Section 8 computer calculations made use of Proposition 3: For green-arrows on a Cartesian system of alternate one-way roads there are no equal length green-arrows with length between 1 and 2. An elementary explanation of why Proposition 3 is valid is given in Section 4.1 and a more general explanation of why this proposition is valid is given in Section A.1.6.

7) For equal length green-arrows it was shown the longest green-arrow on a Cartesian grid of alternate-one-way roads has a length of two and the longest green-arrow length on a Cartesian grid of two-way roads has a length of one (Sections 4, A.1.4, A.1.5, A.2.1).

This result facilitated the development of programs (Appendix B) for finding all proper green-arrow lengths.
8) For equal length green-arrows, green-arrow laws of motion on alternate one-way or two-way roads *cannot* be satisfied for some green-arrow lengths (Section 5).
9) For equal length green-arrows traveling on alternate one-way or two-way roads, green-arrow laws of motion *can* only be satisfied for certain discrete green-arrow lengths (Section 6).
10) For unequal green-arrows traveling on a Cartesian grid of alternate one-way roads, it was shown that the sum of the green-arrow length in the northern/southern direction plus the green-arrow length in the eastern/western have discrete values and that these discrete values are simply related to the discrete proper equal length green-arrow lengths (sections 6 and 7).
11) For anisotropic flow on a Cartesian grid of alternate one-way roads with green-arrows of unequal length, although the sum of the green-arrow lengths traveling in the north/south and east/west directions must sum to a rational number, individual green-arrow lengths may have irrational lengths (Section 7). The observation that proper equal length green-arrows are rational numbers facilitated the development of programs (Appendix B) for finding all proper equal length green-arrow lengths.
12) Discrete values for the sum of unequal length green-arrow lengths in orthogonal directions, for anisotropic flow on alternate one-way streets, are twice the discrete equal length green-arrow lengths on alternate one-way streets (Section 7). Figure 19 b and video-5 suggest this result is also true for two-way streets/roads.
13) Computer code was written which found all useful equal length proper green-arrow lengths for green-arrows traveling on Cartesian alternate one-way or two-way roads (Section 8).

**Applications and discussion**. The most important application of this work is that it enables traffic engineers to turn suburban arterial road networks into maximum throughput uninterrupted flow RGW-road networks. Since vehicles on RGW-roads make every traffic signal provided they follow their green-arrow, RGW-roads are like low speed highways/freeways. When the RGW-roads have equal length green-arrows the throughput of an RGW-road per lane is roughly half the throughput of a highway/freeway. The reason for the roughly one half factor is that, neglecting yellow and all red time, green-arrows on RGW-roads occupy half the area of a lane (video-1, video-2), but on highways/freeways green-arrows occupy the entirety of a lane. Highways/freeways achieved their higher throughput by constructing overpasses/underpasses. Because RGW-roads do not use overpasses/underpasses they are much less expensive to build than highways/freeways. RGW-roads can relieve bumper-to-bumper highway/freeway traffic in suburban areas at a modest cost compared to building additional highways/freeways.

**Conclusions**. Mathematical results given in this paper in sections 3-7, which were derived by algebraic and geometric reasoning, were supported by computer calculations in Section 8 and Appendix B. A more detailed mathematical analysis is given in Appendix A. Mathematical results derived here have already been used [1] to get long progressions on a two-way suburban arterial road with maximum potential throughput. A paper under development will show that the procedures developed here can be used to get arbitrary long progressions with maximum potential throughput on a suburban network of 3X4 intersecting arterials. We believe these procedures can be applied to urban road networks and can lead to more effctive controls and eventually to changes in highway/freeway design .

The development of the RTFD and the formulation of green-arrow laws of motion makes practical a method for obtaining arbitrarily long progressions on alternate one- and two-way RGW suburban arterial road networks and, except for red time and intersection clearing time, each through lane can be utilized by vehicle platoons 100% of the time in the forward direction. The RFTD reduces motorist and bus traffic light stops which encourages motorist, pedestrian and bus travel.   This can provide substantial environmental benefits.

## APPENDIX A – ANALYTICAL DERIVATIONS

This appendix analytically demonstrates the truth of several results suggested by the numerical explorations of Section 8 in a more general setting.  We begin by defining the wavelength concept.

### Definitions of $\lambda$, spatial period for anisotropic flow, $\alpha^+, \alpha^-, \beta^+$ and $\beta^-$.

Symbols $\alpha$ and $\beta$, are green-arrow lengths in the east/west and north/south directions.  The symbol $\lambda$ denotes the wavelength and is defined by

$$\lambda \equiv \alpha + \beta \tag{14}$$

For isotropic flow, $\alpha = \beta$ which implies

$$\lambda_{iso} = 2\alpha = 2\beta \tag{15}$$

Figure 39 describes spatial period and the influence of green-arrow lengths in one direction on the spacing between arrows in the orthogonal direction.

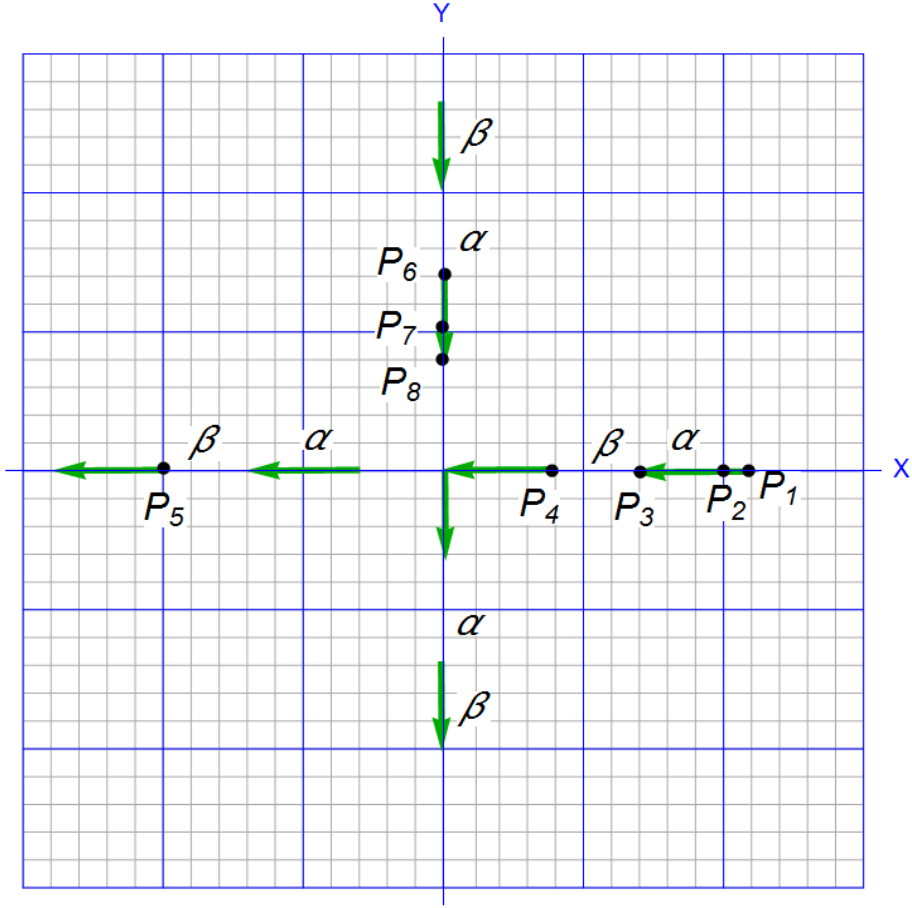

**Figure 39.** Illustration of unequal green-arrow length periodicity on a Cartesian alternate one-way road network.

In the eastern/western directions green-arrow length is denoted by $\alpha$ and the space between green-arrows is denoted by $\beta$. Green-arrow laws of motion then imply that green-arrows traveling in the northern/southern directions have green-arrows of length $\beta$ separated by blank spaces of length $\alpha$. Wavelength $\lambda$ in the east/west direction is the distance between $P_1$ and $P_4$. There are three wavelengths between $P_1$ and $P_5$.

Equations (14) and (15), and the realization that $\mathbb{L}_{EAOW} = \alpha$, (11) is expressed in terms of new variables

$$\frac{4}{\alpha + \beta} = \frac{4}{\lambda} = n, \quad n = 1, 2, 3, \dots \tag{16}$$

Equation (16) captures the principal result from Section 7. The sum of green-arrow lengths in two orthogonal directions have discrete proper values. Equation (11) is for isotropic flow. Equation (16) generalizes (11) to anisotropic flow. For the case of isotropic flow, $\mathbb{L}_{IAOW} = \alpha = \beta$ and (16) reduces to (11).

Next, we introduce definitions of the symbols $\alpha^+$, $\alpha^-$, $\beta^+$ and $\beta^-$ to facilitate our analytical derivations. When a green-arrow is partly through an intersection, the part through the intersection and the part not through the intersection are denoted respectively with a plus or minus superscript. In Figure 39, $\beta^-$ and $\beta^+$ are respectively $d(P_6, P_7)$ and $d(P_7, P_8)$ where the notation $d(P_i, P_j)$ denotes the distance between $P_i$ and $P_j$. Similarly, $\alpha^-$ and $\alpha^+$ are respectively $d(P_1, P_2)$ and $d(P_2, P_3)$. Clearly, $\alpha^- + \alpha^+ = \alpha$ and $\beta^- + \beta^+ = \beta$.

With the above new concepts, the results developed in Sections 7 and 8 can be generalized and summarized in the following two theorems.

**Theorem 1.**
*Necessity:* In a Cartesian alternate *one-way* RGW-road network, the wavelength must satisfy (16).
*Sufficiency:* If the wavelength satisfies (16), then an alternate one-way RGW-road network can be realized on a Cartesian grid.

**Theorem 2**.
*Necessity:* In a Cartesian *two-way* RGW-road network, the wavelength must satisfy (18) given below.
*Sufficiency:* If the wavelength satisfies (18), then a two-way RGW-road network can be realized on the Cartesian grid.

Proposition 1 from Section 8 is a special case of Theorem 1 for alternate one-way RGW-road networks with equal length green-arrows (i.e., isotropic flow with $\alpha = \beta$). Proposition 2 also follows from Theorem 1. Similarly, Proposition 3 is a special case of Theorem 2 for two-way RGW-road networks with equal length green-arrows and Proposition 3 follows from Theorem 2. In Sections A.1 and A.2, we will provide analytical justifications for Theorems 1 and 2, respectively.

*A.1 Cartesian alternate one-way road networks*. Although the focus in this section is on alternate one-way roads on a Cartesian Road network, Section 2.8 indicates it is applicable to road networks that are topologically equivalent to a Cartesian alternate one-way road network.

*A.1.1 Counting wavelengths around a Block*. Counting wavelengths around a Block is useful for deriving Theorem 1. This concept is only defined for Blocks where, going either clockwise or counter clockwise, all green-arrows point in the same direction.

**Definition of the unit Block**. We are free to choose the origin at any intersection on the Cartesian Road grid. As indicated in Figure 40 the convention is adopted that waves on the x- and y-axis respectively travel west and south. Consistent with Section 8, the convention is adopted that at time $t = 0$ the westbound green-arrow tip and the southbound green-arrow tail are at the origin. In Figure 40, the unit Block is defined by coordinates $P_1, P_2, P_5$ and $P_6$. More generally, the unit Block is defined as the smallest geographical area completely enclosed by RGW-roads, e.g., the blue lines in Figure 40 define unit Blocks. With the adopted conventions one can always count the number of green-waves around the unit Block.

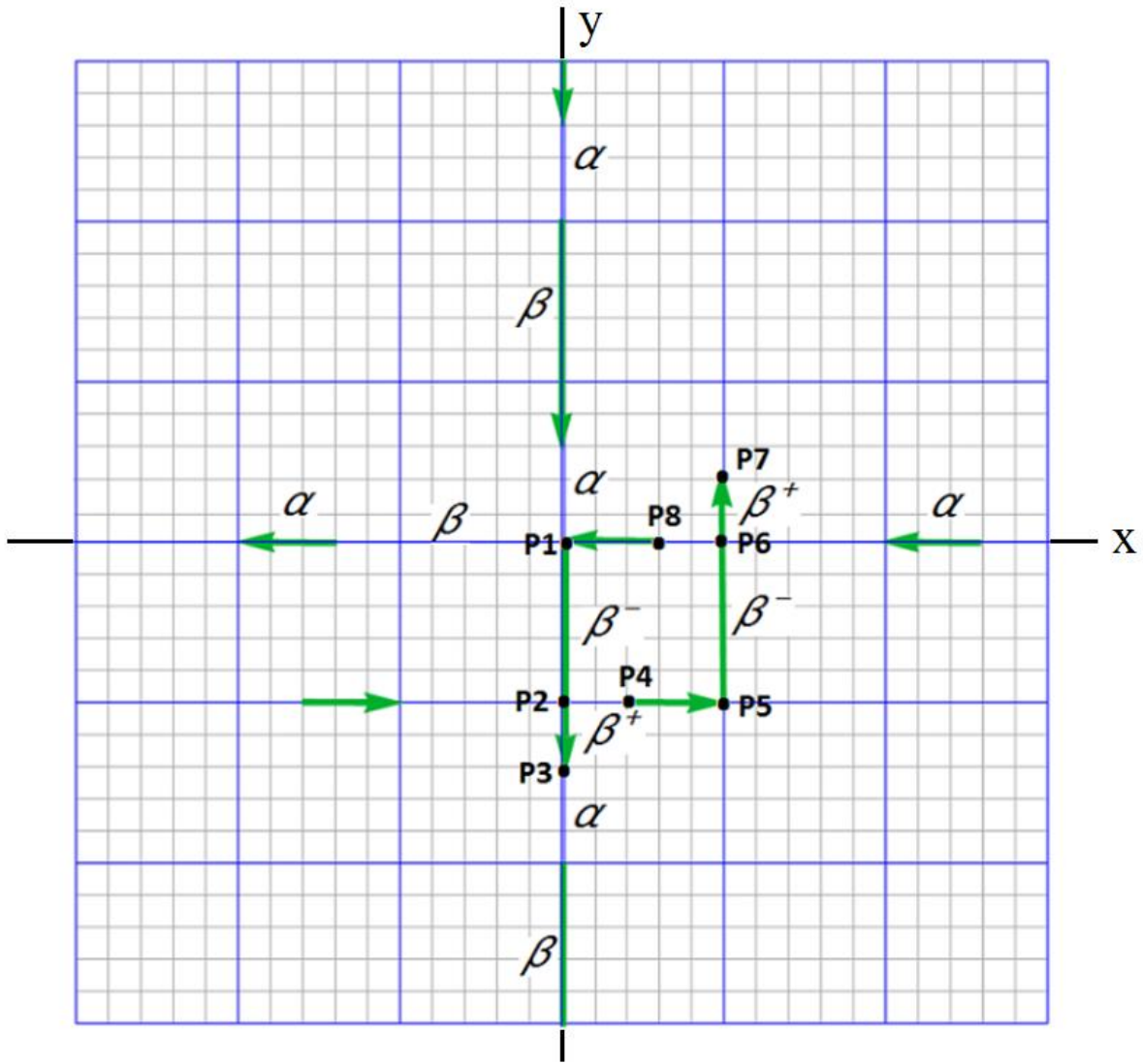

**Figure** 40. Counting waves around a Block for case where $\beta > 1$.

Here $\beta = 7/5$ and $\alpha = 3/5$. The distances $P_1$ to $P_2$ and $P_5$ to $P_6$ equal $\beta^-$ and the distance $P_2$ to $P_3$ and $P_6$ to $P_7$ equal $\beta^+$.

Figure 41 treats the case where $\beta = 4/5 < 1$ and $\alpha = 6/5$. Counting counterclockwise starting at $P_1$, $\beta + \alpha^- + \alpha^+ + \beta + \alpha^- + \alpha^+ = \beta + \alpha + \beta + \alpha = 2\lambda$ we count two wavelengths around the block. Observe the sequence $\beta + \alpha + \beta + \alpha$ encountered in going around the block once corresponds to a mapping from the sides of the block to the sides of the four blocks south of the point $P_1$.

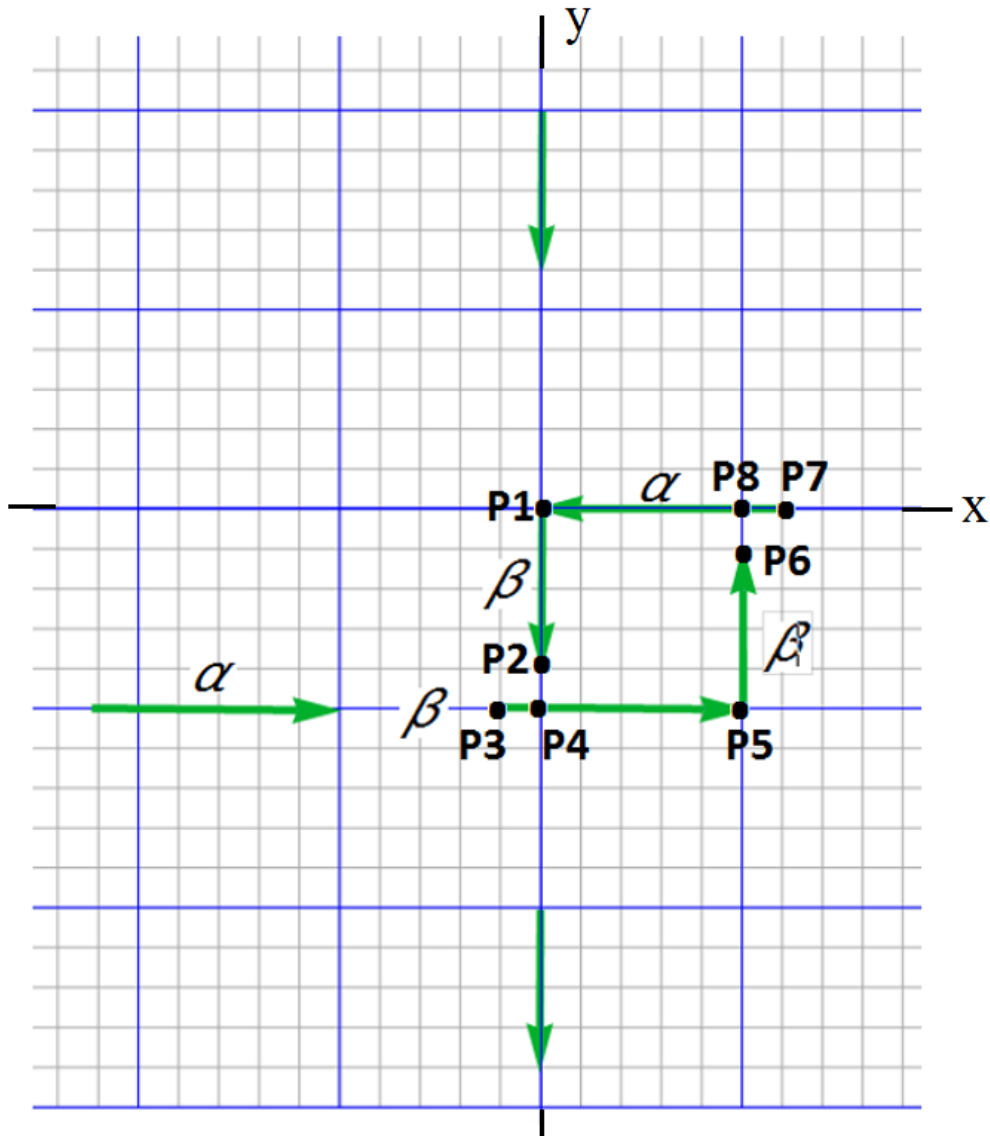

**Figure** 41. Counting waves around a Block for the case $\beta < 1$.

Figure 41 treats the case where $\beta = 4/5 < 1$ and $\alpha = 6/5$. The distance $P_3$ to $P_4$ equals the distance from $P_2$ to $P_4$ equals $\alpha^-$. Similarly, the distance from $P_7$ to $P_8$ equals the distance from $P_6$ to $P_8$, which equals $\alpha^-$. The distance from $P_4$ to $P_5$ equals the distance from $P_8$ to $P_1$, which equals $\alpha^+$.

Counting counter clockwise starting at $P_1$, $\beta + \alpha^- + \alpha^+ + \beta + \alpha^- + \alpha^+ = \beta + \alpha + \beta + \alpha = 2\lambda$ we count two wavelengths around the block. Observe the sequence $\beta + \alpha + \beta + \alpha$ encountered in going around the block once corresponds to a mapping from the sides of the block to the sides of the four blocks south of the point $P_1$.

*A.1.2 Why spatial period of proper green-arrow lengths is 2 or 4.* The objective of this section is to show that Section 2.2 green-arrow laws of motion imply the spatial period of proper green-arrow lengths is either 2 or 4. As shown in Figures 40 and 41, and without loss of generality, a southbound green-arrow is placed, with conventional orientation on a Cartesian road network with its tail coincident with an intersection, which we label as {0,0}. As indicated in Figures 40 and 41, for north or southbound roads the pattern alternates been green-arrow (β-portion) and blank space (α-portion) in the direction of the road according to the properties of a valid green-wave pattern. After crossing four intersections starting from intersection {0,0}, the number of waves (β, α) enumerated, including any fractional part of a wave, is given by $4/\lambda$. Equation 16 implies $4/\lambda$ is $n$, a positive integer, which equals the number of waves contained within the road segment between intersections {0,0} and {0,4}. The α–portion of the last (β, α) wave ends at intersection {0,4}. Applying Eq. 16 again the pattern repeats starting with the $\beta$ portion of the wave. Repeated use of (16) shows the spatial period $\xi = 4$, i.e., every four blocks the pattern repeats.

The spatial period could repeat every 2 blocks and still repeat every 4 blocks. Spatial periods of 1 or 3 are not possible because for these spatial periods the road direction is wrong for a network of alternate one-way roads. Therefore, for proper green-arrow lengths on a Cartesian Road network of alternate one-way roads, the only allowed spatial periods are $\xi = 2$ or 4 in agreement with numerical calculations exhibited in Table 2. This argument proves Proposition 2, which is a special case where $\alpha = \beta$.

*A.1.3 Why green-arrow proper lengths are given by (16).* The objective of this section is to show (16) is necessary using green-arrow laws of motion (Section 2.2), i.e., the necessity part of Theorem 1. The structure of the proof is to show: 1) an integer number of waves around the unit block and 2) the periodic spacing of green-arrows both imply 1) no collisions take place on any of the roads bounding the unit block, 2) the four intersections of the unit block are fully utilized, and 3) that from a unit block other blocks can be constructed on the entire plane.

The necessary part of the proof of Theorem 1 is illustrated in Figure 42 for the case $\alpha = 1$ and $\beta = 1/3$, an anisotropic version of the case listed in Table 2 where $\alpha = \beta = 2/3$. First draw green-arrow labelled 1 at an instant of time when the head of this arrow is passing through the origin. Then green-arrow 2 must be drawn as shown so that flow through the origin has a maximum value. Green-arrow 3 must be drawn as shown so that the head of this arrow arrives at intersection $\{0, -1\}$ just as the tail of green-arrow 2 clears this intersection. Green-arrow 4 must be drawn as shown so that green-arrow 2 enters intersection $\{0, -1\}$ just as arrow 4 leaves this intersection. Green-arrow 6 must be drawn as shown to have maximum flow at intersection {1,0} and green-arrow 5 must be drawn as shown because the spacing between green-arrows in the east and westbound directions must equal the length of green-arrows traveling in the north and southbound directions.

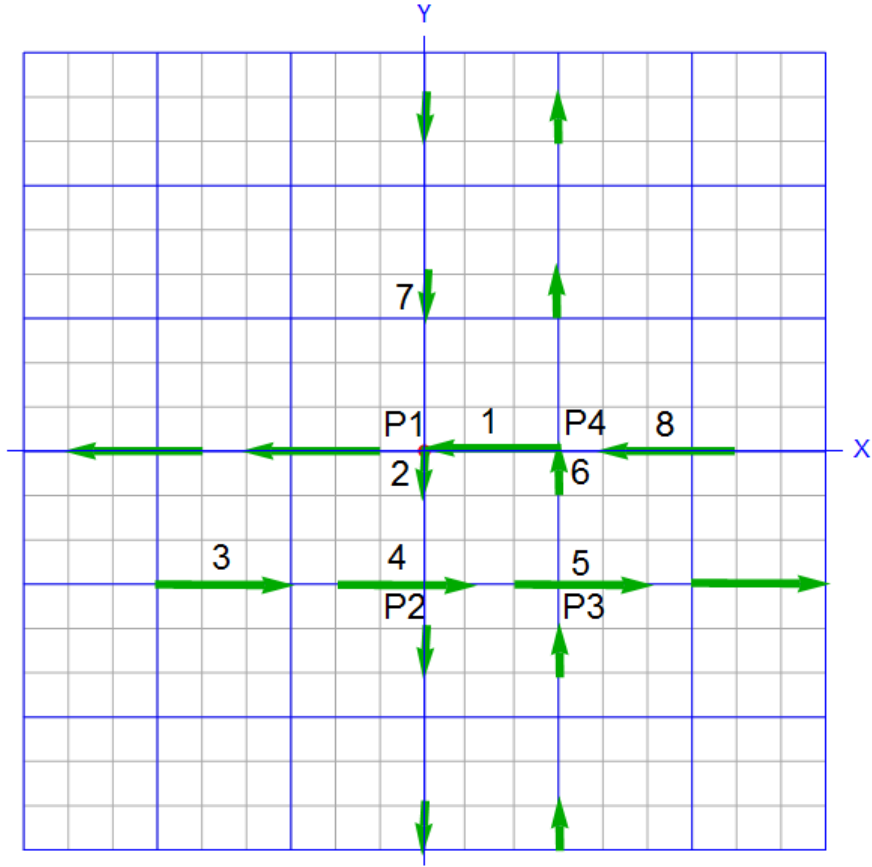

**Figure** 42. Demonstrating: 1) three waves around the unit block $P_1$, $P_2$, $P_3$, $P_4$, 2) maximum flow at intersections $P_1$, $P_2$, $P_3$, $P_4$, and 3) no collisions at intersections $P_1$, $P_2$, $P_3$, $P_4$.

Going counter clockwise, using the method for counting waves around the block documented in Section A.1.1, it is now shown that there are three waves around the unit block. Green-arrow 2 has length $\beta$. The distance from the tip of green-arrow 2 to the tip of arrow 4 is $\alpha$. The distance from the tip of arrow 4 to the tail of arrow 5 is $\beta$. The distance from the tail of arrow 5 to the tail of arrow 6 is $\alpha$. The distance from the tail of arrow 6 to the head of arrow 6 is $\beta$ and the distance from the tail of arrow 1 to the head of arrow 1 is $\alpha$. The total length around the block is $\beta + \alpha + \beta + \alpha + \beta + \alpha = 3\,\lambda$.

Referring to Figure 42 we now show that intersections $P_1, P_2, P_3,$ and $P_4$ have maximum flow and no collisions. This is obvious if one realizes that green-arrows move at the same speed in the east/west and north/south directions. Details are presented for intersection $P_1$. Advance green-arrow 1 three grey squares and do the same for green-arrow 7. At this moment green-arrow 1 is just leaving intersection $P_1$, green-arrow 7 is about to enter that intersection and the head of green-arrow 8 is one grey square east of intersection $P_1$. Advance green-arrow 7 one grey square. At this instant green-arrow 7 is leaving intersection $P_1$ and green-arrow 8 is entering intersection $P_1$. Doing this repeatedly it is clear that intersection $P_1$ is always fully utilized and there are never green-arrow collisions at this intersection. Following this procedure for intersections $P_2, P_3$ and $P_4$ one finds each of these intersections are fully utilized (potential maximum flow) and there are no green-arrow collisions.

We now show how to make new Blocks with integer number of waves circling them at the north-east, north-west, south-east and south-west corners of the unit Block (Blocks B1, B2, B3 and B4). In Figure 43 new-green-arrows $9 - 16$ were added to Figure 42.

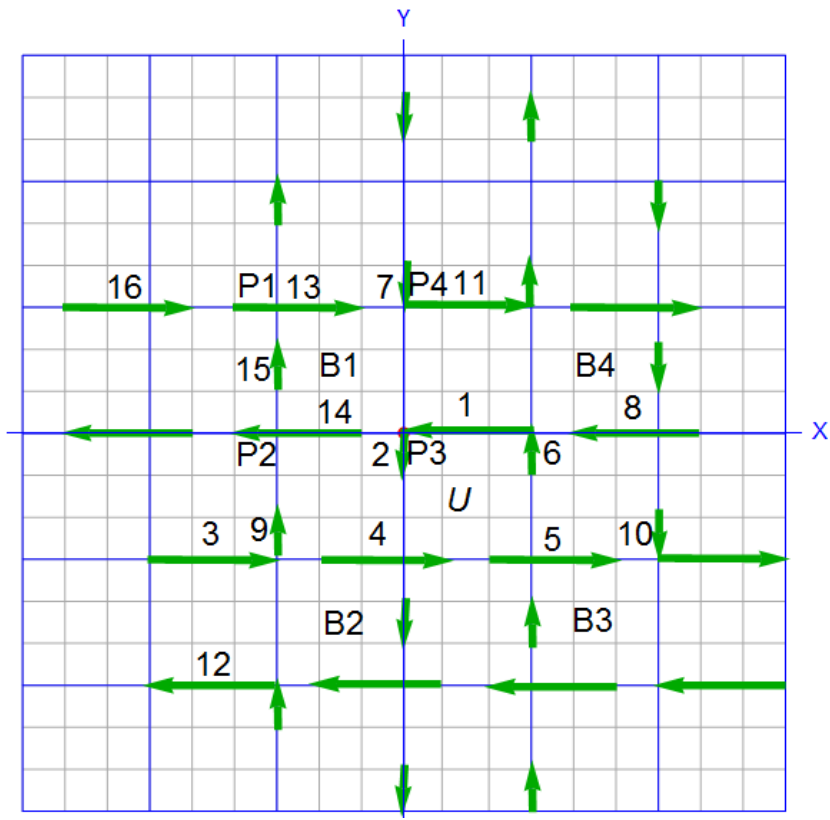

**Figure**. 43. By suitably adding green-arrows to Figure 33, four new blocks $B_1, B_2, B_3$ and $B_4$ have been added to the diagram.

Figure 43 demonstrates: 1) three waves around blocks $B_1, B_2, B_3$ and $B_4$, 2) maximum flow through corners of each new block, and 3) no collisions at corners of each new block.

Waves around $B_1$ are now enumerated using the technique of Section A.1.1. The clockwise distance starting at the tip of green-arrow 15 to the tip of green-arrow 13 equals $\alpha$ and the distance from the tip of green-arrow 13 to the tip of green-arrow 7 equals $\beta$. The distance from the tip of green-arrow 7 to the tail of green-arrow 2 equals $\alpha$ and the distance from the head of green-arrow 1 to the tail of green-arrow 14 equals $\beta$. The distance from the tail of green-arrow 14 to the tail of green-arrow 15 equals $\alpha$ and the distance from the tail of green-arrow 15 to its head equals $\beta$. Adding up all the lengths we find there are three waves around $B_1$. Similarly, it is easy to show there are three waves around blocks $B_2, B_3$ and $B_4$.

It will now be shown that intersection $P_1$ in Figure 43 is fully utilized and that there are no green-arrow collisions at this intersection. After advancing all green-arrows in Figure 43 one grey square the tail of green-arrow 13 and the head of green-arrow 15 are at intersection $P_1$. Advance all arrows in Figure 43 one grey square and the tail of green-arrow 15 is entering the intersection just as the head of green-arrow 16 is entering the intersection. It is apparent that if this is done repeatedly intersection $P_1$ always has maximum flow with no green-arrow collisions. The same argument can be used on intersections $P_2, P_3$ and $P_4$. A similar argument can be made for blocks $B_2, B_3,$ and $B_4$.

Identifying the unit square $U$ as a black square on a checkerboard Blocks $B_1, B_2, B_3,$ and $B_4$ are also black squares on a checkerboard. The procedure outlined in the previous paragraph can be extended to all black squares and these include every intersection on the Cartesian Road network. Equations (11) and (16) were derived from the requirement that the number of waves around the unit block be an integer. Using particular values for $\alpha$ and $\beta$, this section has shown that this requirement also implies each intersection of the two-dimensional grid has maximum flow and there are no green-arrow collisions. Although the proof was done for particular values of $\alpha$ and $\beta$ the proof could have been done for any proper $\alpha$ and $\beta$ given by (16). This establishes the necessity part of Theorem 1, i.e., (16) is a necessary condition to satisfy the green-arrow laws of motion.

*A.1.4 Sufficiency of (16)*. Next, we shall show that the sufficiency part of Theorem 1, i.e., the objective of this section is to show that when (16) is satisfied it is possible to satisfy green-arrow laws of motion over the entire plane. The derivation proceeds in two stages. In stage 1 it is shown the requirement that the number of waves around a Block be an integer for the case $\alpha = \beta = 2/3$ is equivalent to (16). The derivation starts with a special case to establish notation in an unambiguous manner. In stage 2 it is shown that when the number of waves around a Block is an integer, then green-arrow laws of motion are satisfied regardless of the particular $\alpha$ and $\beta$ values.

Consider a unit Block in the grid. Without loss of generality, we assume a green-arrow has its tail at an intersection and is heading south as indicated in Figure 44. Using the procedure described in Figures 40 and 41, start at $P_1$ and count green-arrows and spaces as we make a circuit of a Block. As indicated in Figures 40 and 41 the count will always begin with $\beta$ followed by the $\alpha$ portion. The mandatory maximum flow condition through each intersection implies that when a green=arrow of length $\beta$ leaves an intersection (as in Figures 40 and 41) then a green-arrow of length $\alpha$ enters the intersection and this implies the counter clockwise sequence around a block end in $\alpha$. When there are an integer number of $\{\beta, \alpha\}$ sequences around the block the $\{\beta, \alpha\}$ pattern endlessly repeats.

**Stage 1**. In this stage we show for the special case of $\alpha = \beta = 2/3$ the number of waves around a block is an integer and this implies green-arrow laws of motion (Section 2.2) are satisfied over the entire plane. Figure 44 demonstrates there are 3 waves around the unit block with corners at $P_1, P_2, P_3,$ and $P_4$.

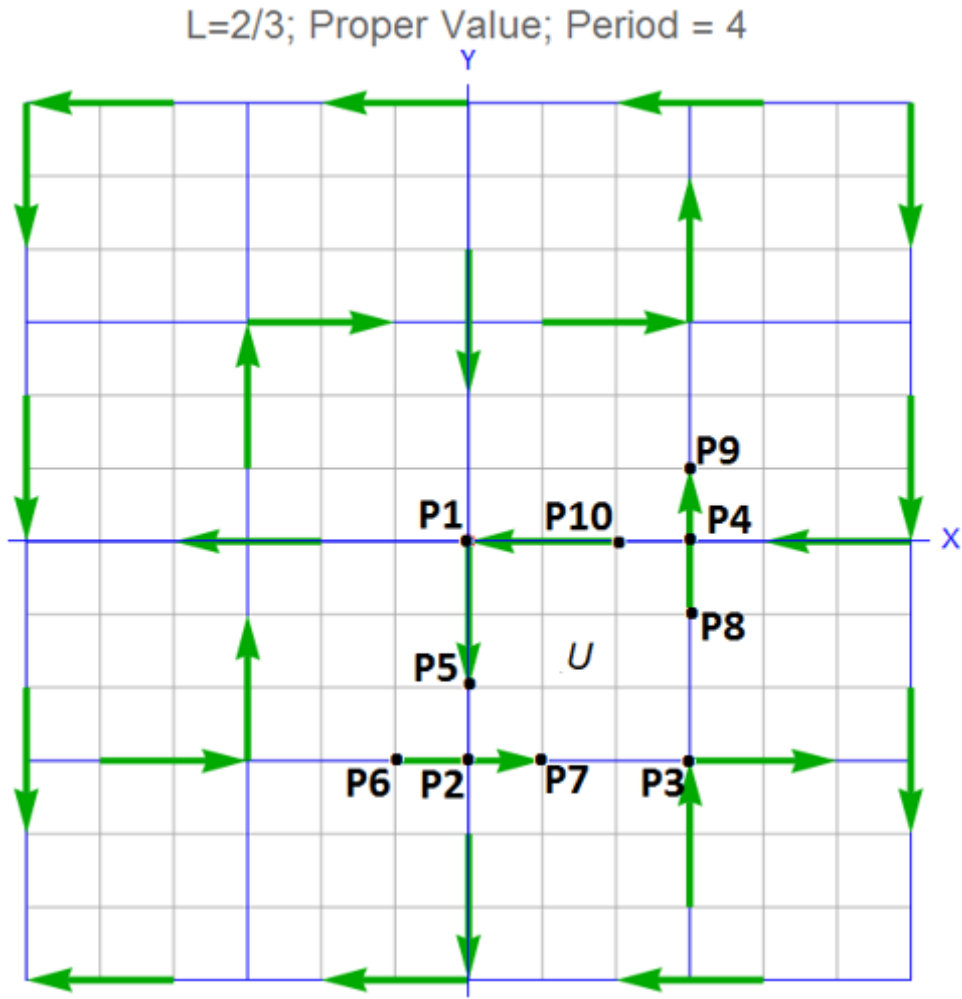

**Figure** 44. Demonstrating integer number of waves around the unit Block with corners $P_1, P_2, P_3$, and $P_4$ for the case where $\alpha = \beta = 2/3$.

The distance from $P_1$ to $P_5$ equals $\beta$ and the distance from $P_5$ to $P_2$ equals the distance from $P_6$ to $P_2$ equals $\alpha^-$. The distance from $P_2$ to $P_7$ equals $\alpha^+$ and the distance from $P_7$ to $P_3$ equals $\beta$. The distance from $P_3$ to $P_8$ equals $\alpha$ and the distance from $P_8$ to $P_4$ equals $\beta^-$. The distance from $P_4$ to $P_{10}$ equals the distance from $P_4$ to $P_9$ equals $\beta^+$ and the distance from $P_{10}$ to $P_1$ equals $\alpha$. Adding up the lengths we get $\beta + \alpha^- + \alpha^+ + \beta + \alpha + \beta^- + \beta^+ + \alpha = 3\lambda$ which shows that Figure 44 has an integer number of wavelengths around the Block.

Figure 45, which is an extended view of Figure 44 demonstrates that over the entire plane there are no collisions (for the case $\alpha = \beta = 2/3$). In this figure Block edges are labeled using a method that will now be described. Going counterclockwise areound a block from the origin green-arrows head south, east, north and west. We arbitrarily label the edges of this block 1S, 2E, 3N and 4W. The word arbitraily is used since we could have used any Block to label Block edges and in the Block chosen we could have started at any corner and gone around the Block in either the clockwise or counterclockwise directions. For example, had we started at the origin and gone around the block in clockwise direction the edges of the square would have been labeled 1W, 2N, 3E and 4S or they could have been labeled 1W, 1N, 1E and 1S.

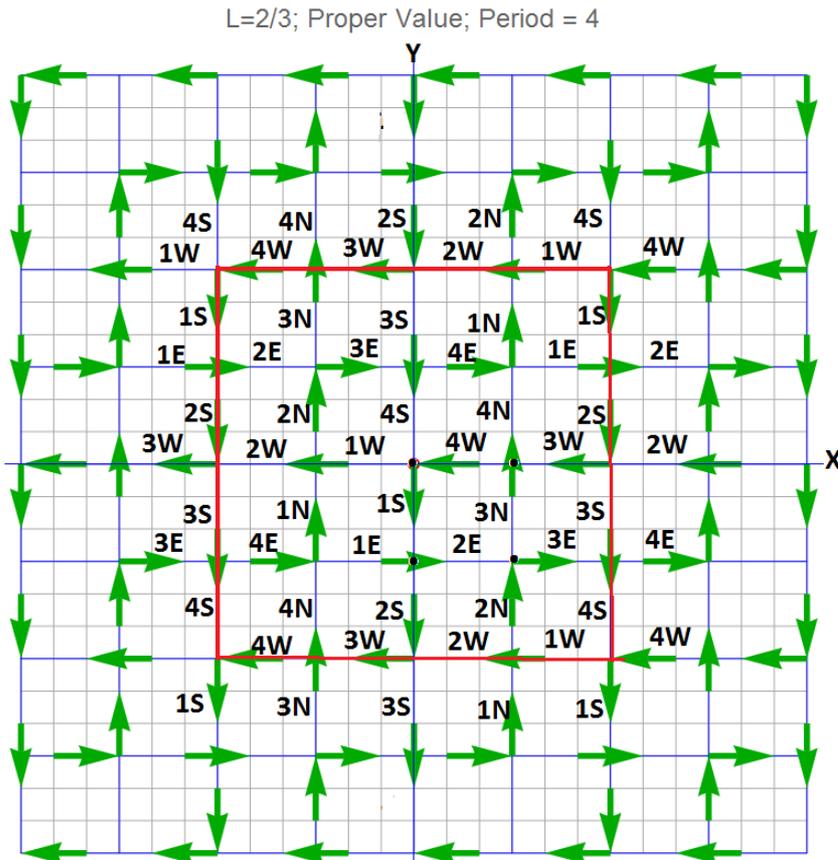

**Figure** 45. Demonstration that with integer number of waves around a Block, green-arrow laws of motion are satisfied. Here $\alpha = \beta = 2/3$. Edges of each Block are designated with a label, for example **1S**.

The red square sorrounding the origin is a basic building Block that can be used to tile the entire $x - y$ plane by translating it repeatedly 4 Blocks east, west, north or south. Within the Block there are 16 distinct edges labeled $1S, 2S, ..., 1N, .., 1W, ... 1E, ... 4E$. which reflects the observation that the spatial period $\xi = 4$ in the north-south and east-west directions. Examination of all intersections within and on the red square shows that each intersection satisfies the green-arrow laws of motion of Section 2.2. Since the entire $x - y$ plane can be tiled with the contents of the red square including its sides (for clarity,subscripts are not used in Figure 36) , this demonstrates green-arrow laws of motion are satisfied on the entire $x - y$ plane for the case $\alpha = \beta = 2/3$.

**Stage 2**. The argument illustrated in Figure 45 is now used to show that an integer number of wavelengths around a Block implies green-arrow laws of motion are satisfied.

The argument is facilitated by adopting some conventions. It is known that every proper green-arrow diagram has a period of 4. As in Figure 45, the convention is that edges are numbered sequentially $1, 2, 3, 4, 1, 2$ ... and have a subscript (to make Figure 36 easier to read subscripts are not used in this figure) $N, S, E,$ or $W$ depending on the green-arrow direction. The convention is also adopted that Block edge numbering start at the origin and procede counter clockwise around the Block till returning to the origin. As illustrated in Figure 45 these conventions uniquely specify a unique lable for each Block edge for every edge in the entire grid regardless of how large the grid is. For example a $X_A$ is the same wherever it occurs for $X = 1, 2, 3,$ or $4$ and $A = N, S, E,$ or $W$. This result is easily confirmed by examination of Figure 45 and is a consequence of our labeling system and the spatial periodicity of 4 established in Section A.1.3.

It is helpful to enumerate the boxes in the red Blocks of Figure 45. Table 5 displays the results.

**Table 5. Enumeration of Blocks in Figure 45**

$\{\{1_S, 2_E, 3_N, 4_W\}, \{2_S, 2_W, 2_N, 2_E\}, \{3_S, 4_E, 1_N, 2_W\}, \{4_S, 4_W, 4_N, 4_E\},$
$\{1_N, 1_E, 1_S, 1_W\}, \{2_N, 1_W, 4_S, 3_E\}, (3_N, 3_E, 3_S, 3_W\}, \{4_N, 3_W, 2_S, 1_E\}\}$

There are 16 Blocks in Figure 45 but only 8 of them are distinct. Each of these Blocks have four distinct intersection types designated here by the letters $\mathbb{A}, \mathbb{B}, \mathbb{C},$ and $\mathbb{D}$. Use bold numbers to indicate a number with a subscript designating its direction

$$\begin{aligned} \mathbf{1} &= \{1_S, 1_N, 1_E, 1_W\} \\ \mathbf{2} &= \{2_S, 2_N, 2_E, 2_W\} \\ \mathbf{3} &= \{3_S, 3_N, 3_E, 3_W\} \\ \mathbf{4} &= \{4_S, 4_N, 4_E, 4_W\} \end{aligned} \qquad (17)$$

In any proper green-arrow diagram there are four types of intersections:
- $\mathbb{A} = \{\mathbf{1}, \mathbf{2}\}$: Two road segments of type 1 oriented toward intersection and two of type 2 oriented away from the intersection.
- $\mathbb{B} = \{\mathbf{2}, \mathbf{3}\}$: Two road segments of type 2 oriented toward intersection and two of type 3 oriented away from the intersection.
- $\mathbb{C} = \{\mathbf{3}, \mathbf{4}\}$: Two road segments of type 3 oriented toward intersection and two of type 4 oriented away from the intersection.
- $\mathbb{D} = \{\mathbf{4}, \mathbf{1}\}$: Two road segments of type 4 oriented toward intersection and two of type 1 oriented away from the intersection.

Figure 46 shows that these intersections occur at the corner of the unit block with a type $\mathbb{D}$ intersecction at its northwest corner and at the origin. By constuction this unit block has maximum flow and no collisions, i.e., the $\mathbb{A}, \mathbb{B}, \mathbb{C},$ and $\mathbb{D}$ intersections on the unit block have maximum flow and no collisions and the same is true for every $\mathbb{A}, \mathbb{B}, \mathbb{C},$ and $\mathbb{D}$ intersection. This shows that providing there is an integer number of wavelengths around the unit block each intersection satisfies the maximum uninterrupted flow and no green-arrow intersection properties of Section 2.2.

**Comment on proof**. Alternatively, as discussed in Section 8, sufficiency of (16) was established numerically for isotropic flow using the software in Appendix B for different values of *n*. Generalization of this numerical exploration for non-isotropic flow ($\alpha \neq \beta$) is trivial. Although the above analytical proof showed there is maximum flow and no green- arrow intersections, it did so at a particular instant of time. The following second proof of the result emphasizes that when there is an integer number of wavelengths around the unit block there are no green-arrow colisions at any instant of time.

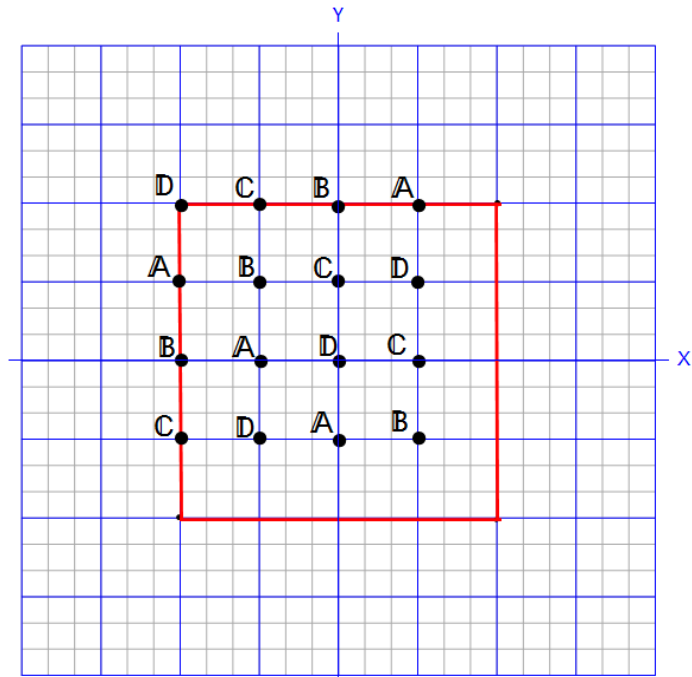

**Figure** 46. Illustrating the location of the four types of intersections.

**Identify conditions when green-arrows never collide**. In Figure 47 each green-arrow was advanced one block and the edges of each block have the same names that were used in Figure 45. We observe that when there is an integer number of waves around a block, green-arrows never collide.

**A conservation law and its proof**. Note the type $\mathbb{A}, \mathbb{B}, \mathbb{C}$, and $\mathbb{D}$ intersections in Figure 46 are cyclic in the horizontal and vertical directions. We claim this cyclic behavior is maintained whenever green-arrows advance one block. Note that letters are in the proper cyclic order in the direction of the road. Thus on westbound roads the correct order is $\mathbb{D}, \mathbb{C}, \mathbb{B}, A$

We now show that this property is conserved whenever each arrow advances one block. Figure 47 illustrates what happens when each green-arrow of Figure 45 advances one block and Figure 48 shows a map of how intersections of type $\mathbb{A}, \mathbb{B}, \mathbb{C}$, and $\mathbb{D}$ transform under this operation. The behavior can be understood in two ways: 1) on northbound or southbound roads each letter advances one block in the direction of motion; 2) on eastbound or westbound roads each letter advances on block in the direction of motion. We observe that either of these transformations leaves the cyclic behavior of the letters $\mathbb{A}, \mathbb{B}, \mathbb{C}$, and $\mathbb{D}$ unchanged.

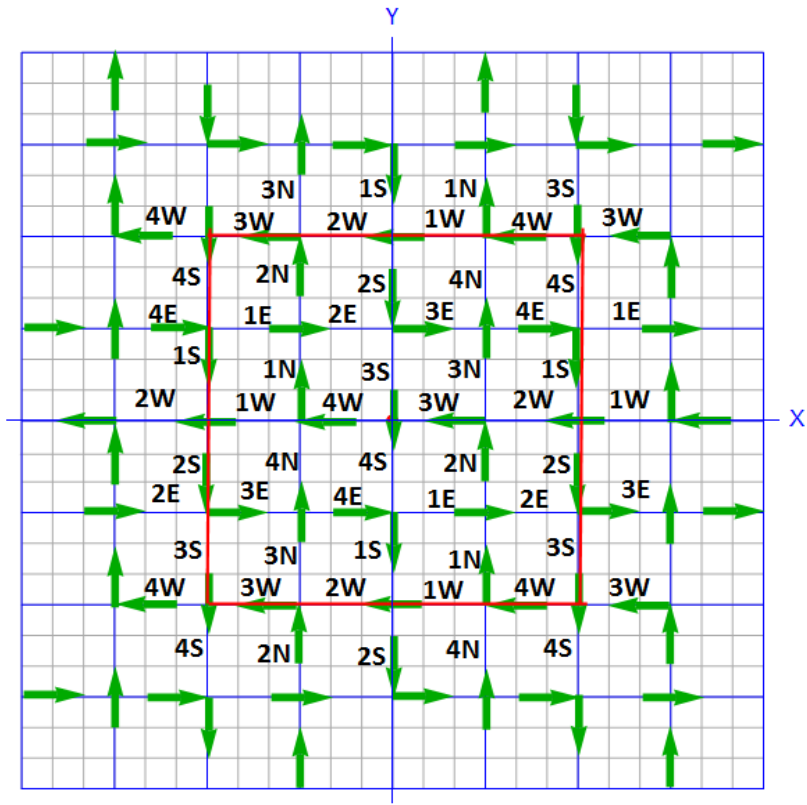

**Figure** 47. Here every arrow in Figure 36 has advanced one Block and the edge of each Block has been named using the same notation used in Figure 36.

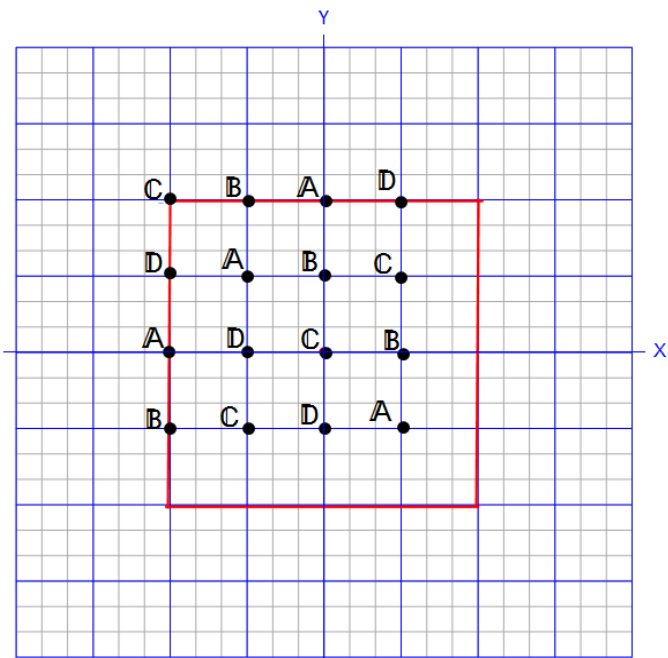

**Figure** 48. Map of intersection types for Figure. 47.

*A.1.5 Isotropic green-arrow properties on a Cartesian grid of alternate one-way roads.* In the isotropic case $\alpha = \beta = \mathbb{L}_{EAOW}$ equation (16) becomes $\mathbb{L}_{EAOW} = 2/n$, $n = 1, 2, ...$ . This implies

- all isotropic proper green-arrow lengths are rational numbers;
- the maximum isotropic green-arrow length is 2;
- there are no proper isotropic green-arrow lengths between 1 and 2.

Equation (11), a special case of (16), shows all proper isotropic green-arrows are rational numbers, 2) the maximum isotropic green-arrow length is two and 3) there are no proper isotropic green-arrow lengths between 1 and 2. The spatial period of the green-arrow pattern in this case is either 2 or 4 because this does not depend on the relative values of $\alpha$ and $\beta$. These results were stated in Section 8 as Propositions 1 and 2.

**Relation to other work**. The observation that green-arrow lengths are rational numbers shows the search done in Appendix B for proper green-arrow lengths did not need to include irrational numbers. The observation that the maximum isotropic green-arrow length is 2 from (11) and (16) confirms this result, found using more elementary means, in Section 4.1.

*A.2 Generalizing isotropic Cartesian two-way road results to anisotropic flow*. Although the results of this section are done on a Cartesian two-way road network, Section 2.8 indicates results found in this section are applicable to anisotropic flow on road networks topologically equivalent to a Cartesian road network.

*A.2.1 Generalizing (12) to anisotropic flow*. The generalization of (12) to anisotropic flow is

$$\lambda = \alpha + \beta = \frac{2}{n}, \quad n = 1, 2, \ldots. \tag{18}$$

For the isotropic case $\alpha + \beta = \mathbb{L}_{ETW}$ which implies (18) is a generalization of (12). We first prove the necessity of (18), stated in Theorem 2. Note that a proper green-arrow pattern (one that satisfies Section 2.2. green-arrow laws of motion) can be decomposed into two separate one-way green-arrow traffic patterns, each going in the opposite direction.

**Demonstrate $\lambda \leq 2$**. It will now be shown that $\lambda \leq 2$. This is a generalization of the result, shown in Section 4.2, that proper isotropic green-arrows have a maximum length of 1. Our approach is to assume that $\lambda > 2$ and show that this assumption leads to a collision of green-arrows. Without loss of generality, choose a moment of time with the following configuration of green-arrows:
  a) one southbound green-arrow of length $\beta$ with tail at the origin,
  b) one northbound green-arrow with tail at the origin,
  c) one westbound green-arrow with head at the origin
  d) one eastbound green-arrow with head at the origin

Figure 49 facilitates a proof that $\lambda$ cannot be greater than 2 and still have a proper green-arrow diagram.

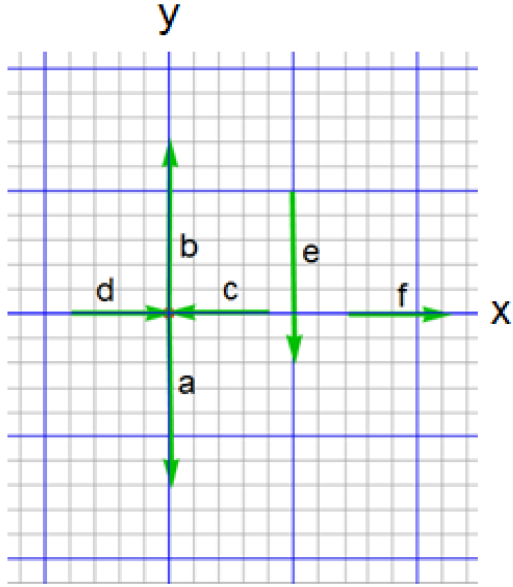

Figure 49. Demonstration that $\lambda$ greater than 2 is not possible for proper anisotropic flow on two-way Cartesian road network. Here $\beta = 7/5$ and $\alpha = 4/5$ so that $\alpha + \beta = 11/5 > 2$.

Arrows a) and c) determine a valid one-way traffic green-arrow pattern denoted by GA1 that covers the entire grid. Similarly, arrows b) and d) determine a valid one-way traffic green-arrow pattern denoted by GA2 with opposite direction to GA1. Without loss of generality assume $\beta > \alpha$. Since $\lambda = \alpha + \beta > 2$, which implies $\beta > 1$ and $\beta - 1 > 1 - \alpha$. Consider green-arrow (d) in GA2. Traversing the x-axis in GA2, arrow (d) is followed by a blank space of length of length $\beta > 1$. This blank space will overshoot the vertical line $x = 1$ by $\beta - 1$. In GA2, there will be a green-arrow, call it green-arrow (e) with tail at $\{1,1\}$ and head at $\{1, -(\beta - 1)\}$. Clearly, if $\alpha > 1$ green-arrow (c) crosses arrow (e). Assume instead that $\alpha \leq 1$. If arrow (c) moves back by $1 - \alpha$ (i.e., eastward) its tail will coincide with the point $\{1,0\}$. The movement of arrow (c) implies arrow (e) also moves back by $1 - \alpha$ units northward so its head will be at the point $\{1, -(\beta - \alpha)\}$. Since $(\beta - \alpha) > 0$, this shows that arrows (c) and (e) collide in this case as well. Therefore, the assumption that $\lambda > 2$ must be false and we have shown that for anisotropic flow on two-way roads $\lambda \leq 2$.

**Demonstrate $\lambda = 2/n$.** Equation (16) asserts that for GA1 and GA2 to be valid $\lambda = 4/n$ with $n = 1, 2, \ldots$. We have shown that for two-way traffic $\lambda \leq 2$. Thus, we conclude for two-way traffic, $\lambda = 2/n$ for $n = 1, 2, \ldots$ which demonstrates the validity of (18).

*A.2.2 Sufficiency of (18).* The sufficiency part of Theorem 2 for two-way RGW-road networks can be shown in a similar way to the sufficiency part of Theorem 1, which was proved in Section A.1.4. In the interest of space, we have omitted the analytical proof. Alternatively, as discussed in Section 8, sufficiency of (18) was established numerically for isotropic flow using the software in Appendix B for different values of $n$. Generalization of this numerical exploration for non-isotropic flow ($\alpha \neq \beta$) is trivial.

# APPENDIX B - SOFTWARE

Links to Mathematica code used in Section 8 of the paper as well as code listings in pdf format are made available here. Mathematica software (player) is needed to run (see) the code.

**Exercising Mathematica code**.
- Wolfram allows one to try https://www.wolfram.com/mathematica/trial/ Mathematica for a limited period of time without charge.
- Wolfram player may be downloaded without charge https://www.wolfram.com/player/

**Alternate one-way roads**.
- A link to a Word file that describes Mathematica code which generates green-arrow patterns for alternate one-way roads.
- A link which allows one to download Mathematica code which generates green-arrow patterns for alternate one-way roads.
- A link to a pdf catalogue of green-arrow patterns for alternate one-way roads.

**Two-way roads**.
- A link to a pdf file that describes Mathematica code which generates green-arrow patterns for two-way roads.
- A link to a pdf file that shows how to use the code for generating green-arrow patterns for two-way roads.
- A link to Mathematica code which generates green-arrow patterns for two-way roads.
- A link to a pdf catalogue of green-arrow patterns on two-way roads.

# Conflicts of Interest

The authors declare that there are no conflicts of interest regarding the publication of this article.

# Funding Statement

No funding was received for this manuscript.

# Acknowledgments

# Supplementary Materials

The supplementary materials include files for video-1, video-2, video-3, video-4, video-5, a one-page summary of invention, and the software-related files listed in Appendix B.